\documentclass[prd,preprint,tightenlines,floatfix,showpacs,preprintnumbers,nofootinbib,eqsecnum]{revtex4}

  \usepackage{amssymb}
   \usepackage{amsmath}
    \usepackage{amsfonts}
     \usepackage{epsfig}
      \usepackage{bm} 
      \usepackage[dvipsnames,usenames]{color}

 \def\Pom{{ I\!\!P}}
 \def\Reg{{ I\!\!R}}
 \def\gsim{\mathrel{\rlap{\lower4pt\hbox{\hskip1pt$\sim$}}
 \raise1pt\hbox{$>$}}}

 \newcommand\la{\langle}
 \newcommand\ra{\rangle}
 \newcommand\beq{\begin{equation}}
 
 \newcommand\eeq{\end{equation}}
 \newcommand\beqn{\begin{eqnarray}}
 \newcommand\eeqn{\end{eqnarray}}

\def\mb{\,\mbox{mb}}

\def\GeV{\,\mbox{GeV}}

\def\Pom{{ I\!\!P}}
\def\Reg{{ I\!\!R}}

\def\lsim{\mathrel{\rlap{\lower4pt\hbox{\hskip1pt$\sim$}}
    \raise1pt\hbox{$<$}}}         
\def\gsim{\mathrel{\rlap{\lower4pt\hbox{\hskip1pt$\sim$}}
    \raise1pt\hbox{$>$}}}         

\def\Im{\,\mbox{Im}\,}
\def\mb{\,\mbox{mb}}

\def\GeV{\,\mbox{GeV}}

\def\beq{\begin{equation}}
\def\eeq{\end{equation}}

\def\beqy{\begin{eqnarray}}
\def\eeqy{\end{eqnarray}}

\begin{document}

\title{Diffractive dijet production: breakdown of  factorization}

\author{B. Z. Kopeliovich$^{1}$}
\email{bzk@mpi-hd.mpg.de}

\author{Roman Pasechnik$^{2,3}$}
\email{Roman.Pasechnik@thep.lu.se}

\author{I. K. Potashnikova$^{1}$}
\email{irina.potashnikova@usm.cl}

\affiliation{
{$^1$\sl 
Departamento de F\'{\i}sica,
Universidad T\'ecnica Federico Santa Mar\'{\i}a;\\
Centro Cient\'ifico-Tecnol\'ogico de Valpara\'{\i}so,
Casilla 110-V, Valpara\'{\i}so, Chile
}\\
{$^2$\sl
Department of Astronomy and Theoretical Physics, Lund
University, SE-223 62 Lund, Sweden
}\\
{$^3$\sl Nuclear Physics Institute ASCR, 25068 \v{R}e\v{z}, 
Czech Republic\vspace{1.0cm}
}}

\begin{abstract}
\vspace{0.5cm}
We analyse the origin of dramatic breakdown of diffractive factorisation, observed in single-diffractive (SD) dijet production in hadronic collisions. One of the sources is the application 
of the results of measurements of the diagonal diffractive DIS to the off-diagonal hadronic diffractive process. The suppression caused by a possibility of inelastic interaction with 
the spectator partons is calculated at the amplitude level, differently from the usual probabilistic description. It turns out, however, that interaction with the spectator partons 
not only suppresses the SD cross section, but also gives rise to the main mechanism of SD dijet production, which is another important source of factorization failure.
Our parameter-free calculations of  SD-to-inclusive cross section ratio, performed in the dipole representation, agrees with the corresponding CDF Tevatron (Run II) 
data at $\sqrt{s}=1.96$ TeV in the relevant kinematic regions. The energy and hard scale dependences demonstrate a trend, opposite to the factorisation-based 
expectations, similarly to the effect observed earlier in diffractive Abelian radiation. 
\end{abstract}


\maketitle

\section{Introduction}
\subsection{Why diffractive factorization fails}
\label{Sec:Intro}

Hadronic diffraction at high energies provides opportunities for a better understanding of an interplay between short- and 
long-range QCD interactions. Diffractive processes, even diffractive DIS at high $Q^2$ \cite{kp}, are typically dominated by 
soft interactions that are difficult to predict from first-principle QCD \cite{Kaidalov:1979jz,Kopeliovich:2006fp}. A special class of 
hard diffractive reactions that necessarily involves a large rapidity gap and hard interactions, in particular  high-$p_T$ particle 
production, have been intensively studied over past two decades. 

Factorisation of short and long distance interactions has been expected to hold for this class of processes in analogy to inclusive 
reactions. It looked natural to assume that  one can measure the PDFs of the Pomeron in the diffractive DIS, and 
assuming their universality, apply the results to hard diffractive processes in hadronic collisions \cite{ing-schlein}. However, 
CDF data \cite{Abe:1997rg} on diffractive dijet  production revealed a dramatic, order of magnitude, breakdown of such 
a diffractive factorization. The mechanism, leading to failure of factorization, is usually related to the presence of spectator 
partons in hadronic collisions. Sometimes it results in an additional suppression factor, called rapidity gap survival probability. 
The diffraction amplitude, however, is a linear combination of elastic amplitudes of different Fock components of the proton, 
which contain rapidity gaps by default. 

Other mechanisms of factorization breaking, related to the multi-gluon Pomeron structure were proposed in \cite{Alves:1996ue,Collins:1992cv,
Yuan:1998ht}. Differently from diffractive DIS, in hadronic collisions the Pomeron can be attached simultaneously to the projectile gluon 
and to the produced parton pair. In other words, the back-to-back high-$p_T$ pair, which has a lifetime substantially shorter than 
the projectile gluon in the incoming hadron, may be produced during the interaction.

A novel mechanism of diffractive factorization breaking was proposed in Refs.~\cite{Kopeliovich:2006tk,Pasechnik:2011nw} 
for the Drell-Yan process, for gauge and Higgs bosons in Refs.~\cite{Pasechnik:2012ac,Pasechnik:2014lga}, and for diffractive 
heavy flavor production in Ref.~\cite{tarasov2}. For a review on breakdown of diffractive factorisation in hadronic collisions, 
see Refs.~\cite{Pasechnik:2015fxa,Kopeliovich:2016rts}.
 
The main reason of non-universality of the diffractive structure functions, measured in DIS, is the principal difference 
between the diagonal and off-diagonal diffractive processes. Diffractive DIS, $\gamma^*+p\to X+p$, is predominantly 
diagonal (elastic $\bar qq+p\to\bar qq+p$), so one should not apply the results of such measurements to the off-diagonal 
diffractive processes (e.g. dijets) in hadronic collisions. In terms of the Regge approach diffraction is related to the Pomeron 
exchange, but the Pomerons in the above two cases are different. Even if the Pomeron were a true Regge pole with a universal 
intercept, the residue functions would have very different features, leading to a breakdown of factorization (see below). 
 
Diagonal diffraction, i.e. elastic scattering, with the forward amplitude related via the unitarity relation to the total cross section, 
in terms of the optical analogy can be treated as a shadow of inelastic processes. The stronger are inelastic interactions, the larger 
is the elastic cross section. The maximum is reached at the unitarity bound, so called `black-disk'' limit.

The off-diagonal diffractive dynamics is more involved. Extending the optical analogy, one can interpret the off-diagonal diffractive 
amplitude as a linear combination of shadows of different inelastic channels, which tend to compensate each others. In the black-disk 
limit they cancel completely, and diffraction vanishes. These features follow from the quantum-mechanical picture of diffraction 
\cite{Glauber:1955qq,fain-pom,Good:1960ba}, which can be illustrated by switching to the eigenstate representation \cite{kl,Kopeliovich:1981pz}.

As far as a hadron is subject to diffractive excitation,  it is apparently not an eigenstate of interaction, but can be expanded over 
the complete set of eigenstates $|\alpha\ra$ of the elastic amplitude operator,  $\hat f_{el}|\alpha\ra =f_\alpha\,|\alpha\ra$ 
\cite{kl,Kopeliovich:1999am,Kopeliovich:2006fp},
\beq
|h\ra = \sum\limits_{\alpha=1}C^h_{\alpha}\,|\alpha\ra\ ,
\label{700}
 \eeq
 where the
coefficients $C^h_{\alpha}$ satisfy the orthogonality  relation,
\beq
\la h'|h\ra  =
\sum\limits_{\alpha=1}(C^{h'}_{\alpha})^*C^h_{\alpha} =
\delta_{hh'}
\label{800}\\
 \eeq
Correspondingly, the elastic and single diffraction hadronic amplitudes can be expressed via 
the eigenamplitudes as,
 \beqn
f_{el}^{h\to h} &=& \sum\limits_{\alpha=1}|C^h_{\alpha}|^2\,f_\alpha
\label{900a}\\
f_{sd}^{h\to h'} &=&
\sum\limits_{\alpha=1}(C^{h'}_{\alpha})^*C^h_{\alpha}\,f_\alpha
\label{900b}
 \eeqn
At the unitarity bound, all the eigen amplitudes $\Im f_\alpha=1$, so the positively defined elastic amplitude (\ref{900a}), as mentioned above, 
reaches a maximum. At the same time, the off-diagonal diffractive amplitude  (\ref{900b}) consists of terms with alternating signs, which 
tend to cancel each other, and the amplitude vanishes in the black-disk limit, according to the orthogonality relation (\ref{800}) 
\cite{Glauber:1955qq,fain-pom,Good:1960ba}.

Frequently, the failure of the predictions based on factorisation, is explained and attempted to be improved by introducing a suppression factor, 
so called gap survival probability, evaluated within probabilistic models  \cite{Ryskin:2009tk,Ryskin:2011qe}. Such an ad hoc way to cure the factorisation 
prescription cannot replace the quantum-mechanical expression (\ref{900b}), so it cannot be correct. The diffractive amplitude (\ref{900b}), is a linear 
combination of elastic amplitudes, which contain a rapidity gap by definition. Therefore, this expression does not need any gap survival factor. 
 
 \subsection{Dipole representation}
 
The eigenstates  of interaction $|\alpha\ra$ in high-energy QCD are color dipoles \cite{Kopeliovich:1981pz}. The eigen amplitudes $f_\alpha$ cannot be 
calculated reliably, but can be extracted from low-$x$ DIS data. Relying on such a color-dipole phenomenology we calculate below the diffractive amplitude 
(\ref{900b}) for dijet production. This process at the Tevatron $p\bar p \to \bar p + {\rm gap} + jj + X$ is characterised by the presence of two jets in the final state, 
a large rapidity gap void of particles, and a leading anti-proton $\bar p$, which survives the collision and remains intact. 

The breakdown of diffractive factorisation, the most striking result of Ref.~\cite{Abe:1997rg}, was seen as an order of magnitude suppression of the measured 
dijet diffractive cross section compared to the theoretical predictions based upon the diffractive parton densities fitted to HERA data on diffractive DIS. 
The main source of this problem, as demonstrated above, is application of the results of the analysis of data on diagonal DIS diffraction to the essentially 
off-diagonal diffractive excitation of hadrons. 

Diffractive gluon Bremsstrahlung off a projectile quark has been studied in the color dipole approach in the limit of small gluon fractional light-cone momentum 
$\alpha\ll 1$ in Ref.~\cite{Kopeliovich:1999am}. In the hadronic case diffractive gluon Bremsstrahlung appears to be the leading-twist process due to interaction 
with the spectator partons \cite{tarasov2}, that is similar to the Abelian case (see e.g. Refs.~\cite{Kopeliovich:2006tk,Pasechnik:2011nw}). While for the forward 
scattering the corresponding process does not vanish (contrary to the Abelian case), QCD factorisation is still expected to be broken due to an interplay between 
hard and soft fluctuations. In this paper, being motivated by the Tevatron data on SD production of dijets, we extend the dipole formalism of 
Ref.~\cite{Kopeliovich:1999am} to the case of arbitrary $\alpha$ of diffractively produced gluon, then we apply it for the hadronic case where large distances 
are necessarily involved and present the key features of the SD-to-inclusive ratio that indicate the dramatic breakdown of diffractive factorisation in non-Abelian diffraction.
The light-cone dipole approach enables to incorporate such effects coherently at the amplitude level, which has been previously proven to work well in the diffractive 
Abelian radiation processes \cite{Pasechnik:2011nw,Pasechnik:2012ac,Pasechnik:2014lga} and diffractive heavy flavor production \cite{tarasov2}. In this paper, following 
the original studies of inclusive \cite{kovch-Mueller,Kopeliovich:1998nw} and diffractive diffractive gluon radiation \cite{Kopeliovich:1999am,tarasov2,Goncalves:2016qku},
we apply the light-cone dipole approach to the analysis  of inclusive and diffractive gluon radiation beyond QCD factorisation. By comparing the dipole model results 
with the Tevatron data for the SD-to-inclusive ratio, we check whether the gap survival effects are properly accounted for in the dipole treatment of the diffractive 
non-Abelian radiation.

The paper is organised as follows. In Section~\ref{Sec:jj-incl}, we develop the dipole model formulation of the inclusive dijet production in the target rest frame 
based upon the gluon Bremsstrahlung mechanism (quark excitation) as well as from the gluon splitting mechanism (gluon excitation). In Section~\ref{Sec:dipole-CS}, 
the models for the universal dipole cross section are briefly discussed in the soft and hard dipole scattering regimes. In Section~\ref{Sec:jj-SD}, we extend the dipole 
formulation to the SD dijet production and derive the corresponding parton- and hadron-level amplitudes as well as the SD cross sections in the hard scattering limit. 
Then, in Section~\ref{Sec:SD-to-incl} we construct the SD-to-inclusive ratio of the cross sections taking into account the CDF Run II experimental constraints on 
the phase space and present the numerical results. Finally, concluding remarks are given in Section~\ref{Sec:summary}.

\section{Inclusive back-to-back dijets}
\label{Sec:jj-incl}

\subsection{Dijets from quark excitations}
\label{Sec:qG-incl}

\begin{figure*}[!h]
 \centerline{\includegraphics[width=0.85\textwidth]{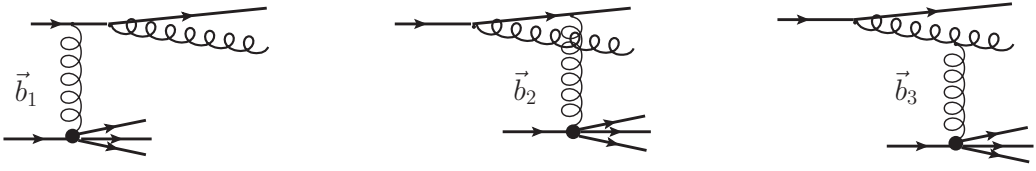}}
   \caption{
   \small The leading-order contributions to the gluon Bremsstrahlung mechanism of 
   high-$p_T$ back-to-back dijets production in quark-nucleon $qN\to qGX$ scattering.}
 \label{fig:hard-dijets}
\end{figure*}

At forward rapidities inclusive production of high-$p_T$ jets in the dipole picture is dominated by the gluon Bremsstrahlung 
mechanism off a projectile quark \cite{Kopeliovich:1998nw} (similar to the Drell-Yan process \cite{Kopeliovich:1995an,Brodsky:1996nj,Kopeliovich:2000fb,Basso:2015pba}). 
The leading order (``skeleton'') diagrams of this process are depicted in Fig.~\ref{fig:hard-dijets}. In this case, $x_1\equiv p^+/P_1^+\lesssim 1$, 
$x_2\equiv p^-/P^-_2\ll 1$, where $p$ is the 4-momentum of the radiated gluon, and $P_{1,2}$ are the 4-momenta of the projectile 
and target nucleons, respectively.

Let us denote the transverse momenta (relative to the projectile quark) of the final quark and gluon as 
$\vec{p}_2$ and $\vec{p}$, respectively, their total momentum as $\vec{q_\perp}=\vec{p}_2+\vec{p}$, and the relative 
momentum as $\vec\kappa = \alpha \vec{p}_2 - \bar\alpha\vec{p}$ in terms of the light-cone momentum fraction $\alpha$ carried 
by the gluon. In the case of collinear projectile quark, the transverse momentum transfer is  equal to $\vec{q_\perp}$. 
Then, the inclusive dijet production amplitude $\hat{B}_{l}({\vec q}_\perp,\vec\kappa)$ reads
\begin{eqnarray}
\hat{B}_{l}({\vec q}_\perp,\vec\kappa) = \int d^2b d^2r e^{i\vec b \vec q_\perp}e^{i\vec r \vec\kappa}\hat{A}_{l}(qN\to qGN_8^*) \,,
\end{eqnarray}
in terms of the corresponding amplitude $\hat{A}_{l}(qN\to qGN_8^*)$ found in impact parameter representation
as sum over three contributions in Fig.~\ref{fig:hard-dijets}
\begin{eqnarray}
\hat{A}_{l}(qN\to qGN_8^*)&=&\frac{\sqrt{3}}{2}\sum_{a} \Big[\tau_{l}\tau_{a}\langle N_8^*|\hat{\gamma}_a(\vec b_1)| N \rangle - 
\tau_a\tau_{l}\langle N_8^*|\hat{\gamma}_a(\vec b_2)| N \rangle \nonumber \\ 
&-&\sum_{b} if_{lab}\tau_b\langle N_8^*|\hat{\gamma}_a(\vec b_3)| N \rangle\Big]\, 
\Psi_{q\to qG}(\vec r,\alpha) \,, 
\label{qG-incl-A}
\end{eqnarray}
where $l$ is the color index of the radiated gluon $G$, $N_8^*$ is the color-octet remnant of the target nucleon, for which 
the completeness relation holds, $|N^*_8 \rangle \langle N^*_8 | = 1$; $\lambda_a=2\tau_a$ are the Gell-Mann matrices;
and $\hat{\gamma}_a$ is the effective gluon-nucleon interaction vertex $GN\to N^*_8$. The impact parameters of the projectile, 
ejectile quarks and the radiated gluon are 
\begin{eqnarray}
\vec{b}_1\equiv \vec{b}\,, \qquad \vec{b}_2\equiv \vec{b} - \alpha\vec r \,, \qquad 
\vec{b}_3\equiv \vec{b} + \bar\alpha\vec r \,, \qquad \bar\alpha \equiv 1-\alpha \,,
\label{imp}
\end{eqnarray}
such that $\vec{r}$ is the transverse separation of the $qG$ system, and $\vec{b}$ is the distance between its center of gravity and the target $N$. 
The light-cone distribution  function for the $qG$ Fock state (with transversely polarised gluon) in the projectile quark $\hat{\Psi}_{q\to qG}$ is given by 
\cite{Kopeliovich:1995an,Brodsky:1996nj,Kopeliovich:1998nw}
\begin{eqnarray} 
\nonumber
&& \hat{\Psi}_{q\to qG}(\vec r,\alpha)=\frac{2}{\sqrt{3}}\frac{\sqrt{\alpha_s}}{2\pi}\chi_f^\dagger \hat{\Gamma} \chi_i\, 
K_0(\tau\,r) \,, \qquad \tau^2=\alpha^2m_q^2+(1-\alpha)m_G^2 \,,  \\
&& \hat{\Gamma}=im_q\alpha^2\, \vec{e}_G\cdot[\vec{n}\times \vec{\sigma}]+
\alpha\, \vec{e}_G\cdot[\vec{\sigma}\times \vec{\nabla}] - i(2-\alpha) \,\vec{e}_G\cdot\vec{\nabla} \,,
\label{psi-qG}
\end{eqnarray}
where $\alpha_s=\alpha_s(\mu^2)$ is the QCD coupling constant determined at the hard scale $\mu^2$, $m_G$ ($m_q$) is the effective gluon (quark) mass;
$\chi$ is the quark spinor, $\vec{e}_G$ is the transverse polarisation vector of the radiated gluon; $K_0(x)$ is the modified Bessel function of 
the second kind; and $\vec{\nabla}\equiv \partial/\partial \vec r$. The corresponding wave function in momentum representation reads
\begin{eqnarray} 
\nonumber
&& \hat{\tilde{\Psi}}_{q\to qG}(\vec \kappa,\alpha)=\frac{2\sqrt{\alpha_s}}{\sqrt{3}}\chi_f^\dagger \hat{\tilde{\Gamma}} \chi_i\, \frac{1}{\kappa^2 + \tau^2} \\
&& \hat{\tilde{\Gamma}}= im_q\alpha^2\, \vec{e}_G\cdot[\vec{n}\times \vec{\sigma}] + i\alpha \vec{e}_G\cdot[\vec{\sigma}\times \vec{\kappa}] - (2-\alpha) (\vec{e}_{G}\cdot\vec{\kappa})\,,
\label{psi-qG-k}
\end{eqnarray}
If the gluon is radiated with large transverse momentum, it is likely to turn into a hard jet, due to intensive radiation.

The differential cross section for the inclusive $qN\to qGX$ process has the form,
\begin{eqnarray}
\frac{d^3\sigma_{\rm incl}(qN\to qGX)}{d(\ln\alpha)d^2\kappa} = \frac13\,\frac{1}{(2\pi)^2}\int \frac{d^2q_\perp}{(2\pi)^2}
\sum_{l}{\rm Tr}\,\Big[\hat{B}^\dagger_{l}({\vec q}_\perp,\vec\kappa)\hat{B}_{l}({\vec q}_\perp,\vec\kappa)\Big] \,,
\end{eqnarray}
where the numerical prefactor indicates at the averaging over colors of the projectile quark. We employ  completeness of 
the remnant $N^*_8$ states and average over the target nucleon degrees of freedom 
as follows,
\begin{eqnarray} 
\label{averaging}
\langle N |\hat{\gamma}_a(\vec b_k)\hat{\gamma}_{a'}(\vec b_l)| N 
\rangle =\frac{3}{4} \delta_{aa'} \phi(\vec b_k,\vec b_l)\,, \qquad \hat{\gamma}_a = \hat{\gamma}_a^\dagger \,.
\end{eqnarray}
Then integrating over ${\vec q}_\perp$ 
one arrives at the differential cross section expressed in terms of the symmetric partial dipole amplitude
$\phi(\vec b_k,\vec b_l)=\phi(\vec b_l,\vec b_k)$,
\begin{eqnarray} \nonumber
\frac{d^3\sigma_{\rm incl}(qN\to qGX)}{d(\ln\alpha)d^2\kappa} = \frac{1}{(2\pi)^2}
\int d^2rd^2r'\, e^{i(\vec r - {\vec r}\,')\vec\kappa} \overline{\sum}\hat{\Psi}_{q\to qG}(\vec r,\alpha) 
\hat{\Psi}^\dagger_{q\to qG}({\vec r}\,',\alpha)\Sigma^{q\to qG}_{\rm eff}(\vec r,{\vec r}\,',\alpha). \\
\label{incl}
\end{eqnarray}
Here the distribution function squared (averaged over the projectile quark spins) is given by
\begin{eqnarray} \nonumber
\overline{\sum}\hat{\Psi} \hat{\Psi}^\dagger &\equiv& 
\sum_{\lambda_g=\pm1}\frac12 \sum_{\sigma_f,\sigma_i}\hat{\Psi}_{q\to qG}(\vec r,\alpha) 
\hat{\Psi}^\dagger_{q\to qG}({\vec r}\,',\alpha) \\
&=&\frac{2\alpha_s}{3\pi^2}\,\Big( m_q^2\, \alpha^4\, K_0(\tau\,r)K_0(\tau\,r') 
+ \Big[1+\bar\alpha^2\Big]\tau^2\, \frac{\vec{r}\cdot \vec{r}\,'}{r r'}\,
 K_1(\tau\,r)K_1(\tau\,r') \Big) \,,
\end{eqnarray}
and the effective dipole cross section reads,
\begin{eqnarray} \nonumber
\Sigma^{q\to qG}_{\rm eff}({\vec r},{\vec r}\,',\alpha)
&=& \int d^2b\; \Big\{
\phi({\vec b}_1,{\vec b}'_1)+\frac18\phi({\vec b}_1,{\vec b}'_2)-\frac98\phi({\vec b}_1,{\vec b}'_3) \\
&+& 
\frac18\phi({\vec b}_2,{\vec b}'_1)+\phi({\vec b}_2,{\vec b}'_2)-\frac98\phi({\vec b}_2,{\vec b}'_3)  \nonumber \\
&-& 
\frac98\phi({\vec b}_3,{\vec b}'_1)-\frac98\phi({\vec b}_3,{\vec b}'_2)+\frac94\phi({\vec b}_3,{\vec b}'_3)\Big\} \,.
\label{eff-dip-ampl}
\end{eqnarray}
It depends on impact parameters,
\begin{eqnarray}
\vec{b}'_1=\vec{b}_1\equiv \vec{b}\,, \qquad \vec{b}'_2\equiv \vec{b} - \alpha\vec r\,' \,, \qquad 
\vec{b}'_3\equiv \vec{b} + \bar\alpha\vec r\,' \,.
\end{eqnarray}

The partial dipole amplitude $\phi(\vec b_k,\vec b_l)$ introduced in Eq.~(\ref{averaging}) is directly related to the universal 
dipole-nucleon cross section $\sigma_{q\bar q}$ as follows (see also Refs.~\cite{Kopeliovich:1998nw,Kopeliovich:1999am})
\begin{eqnarray}
\label{dip-CS-phi}
\sigma_{\bar q q}(\vec{r}_1-\vec{r}_2)\equiv \int d^2 b \Big[ \phi(\vec b+\vec{r}_1,\vec b+\vec{r}_1)+
\phi(\vec b+\vec{r}_2,\vec b+\vec{r}_2)-2\phi(\vec b+\vec{r}_1,\vec b+\vec{r}_2)  \Big]\,,
\end{eqnarray}
so that the $b$-integration in Eq.~(\ref{eff-dip-ampl}) yields
\begin{eqnarray} \nonumber
\Sigma^{q\to qG}_{\rm eff}(\vec r,{\vec r}\,',\alpha)&=&\frac{1}{2}\Big\{
\sigma_{Gq\bar q}\big(\bar\alpha\vec r,\bar\alpha \vec r+\alpha \vec r\,'\big) + 
\sigma_{Gq\bar q}\big(\bar\alpha\vec r\,',\bar\alpha \vec r\,'+\alpha \vec r\big) \\ 
&-&\sigma_{q\bar q}\big(\alpha(\vec r - \vec r\,')\big) - 
\sigma_{\rm GG}\big(\bar\alpha(\vec r - \vec r\,')\big)   \Big\} \,.
\label{eff-dip-CS}
\end{eqnarray}
The gluonic $GG$ dipole cross section \cite{Kopeliovich:2002yv} and the effective three-body $Gq\bar q$ dipole 
cross section \cite{Nikolaev:1994de,Nikolaev:1995ty}, read,
\begin{eqnarray}
\sigma_{\rm GG}(\vec r)= \frac{9}{4}\sigma_{q\bar q}(\vec r) \,, \qquad 
\sigma_{Gq\bar q}(\vec r_1,\vec r_2)= \frac98 \Big(\sigma_{q\bar q}(\vec r_1)+
\sigma_{q\bar q}(\vec r_2)\Big)-\frac18\sigma_{q\bar q}(\vec r_1-\vec r_2) \,,
\label{GG-Gqq}
\end{eqnarray}
respectively. 

In the collinear approximation for the projectile parton, the inclusive hadronic $NN\to qG+X$ cross section reads
\begin{eqnarray} 
 \frac{d^4\sigma^{\rm NN}_{\rm incl}}{d(\ln x_q)\,d(\ln\alpha)d^2\kappa} &=& 
 Q(x_q,\mu^2)\,\frac{d^3\sigma(qN \to qG + X)}{d(\ln\alpha)d^2\kappa} \,, 
 \label{dipole-f}
\end{eqnarray}
where $x_q$ is the fractional light-cone momentum carried by the projectile quark in the parent nucleon, and 
the projectile quark distribution distribution function is
\begin{eqnarray}
Q(x_q,\mu^2)\equiv x_q\,q(x_q,\mu^2)
\end{eqnarray}
at the hard scale $\mu^2$ being the invariant mass squared of the produced $qG$ (or dijet) 
system $\mu^2\simeq M_{qG}^2$.

\subsection{Dijets from gluon excitations}
\label{Sec:GG/qq-incl}

\begin{figure*}[!h]
 \centerline{\includegraphics[width=0.85\textwidth]{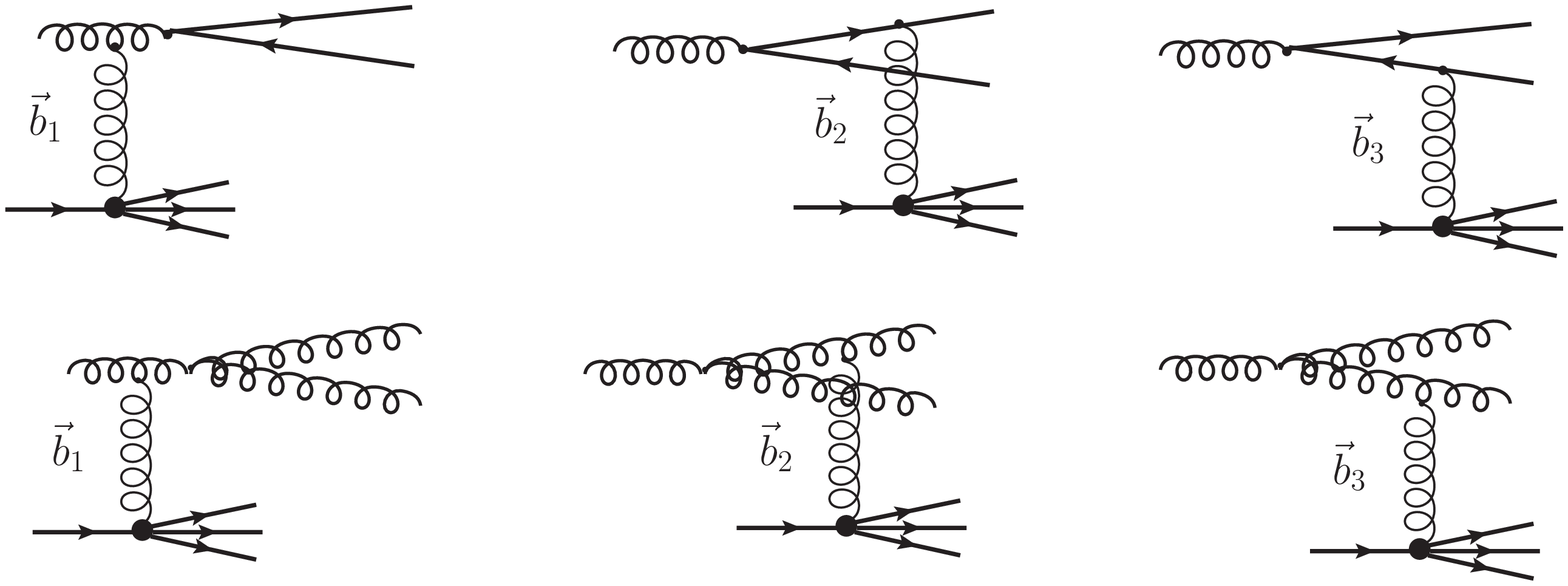}}
   \caption{
   \small The leading-order contributions to high-$p_T$ dijet production
   in gluon-nucleon scattering ($GN\to q\bar qX$ -- upper row, and $GN\to GGX$ -- lower row) 
   in the dipole picture.}
 \label{fig:dijets-GG/qq}
\end{figure*}

At central rapidities inclusive high-$p_T$ dijet production can acquire large contributions from 
the gluon-initiated subprocesses $GN\to q\bar qX$ or $GN\to GGX$, as is shown in Fig.~\ref{fig:dijets-GG/qq} 
by upper and lower rows, respectively.

The amplitude of the inclusive process $GN\to q\bar q N_8^*$ is given by 
the sum of three terms corresponding to the diagrams shown in the upper row of Fig.~\ref{fig:dijets-GG/qq},
\begin{eqnarray}
\nonumber
\hat{A}_{l}(GN\to q\bar q N_8^*) &=& \sqrt{2}\,\sum_{a} 
{\chi_q^\mu}^\dagger  \Big\{\tau_a\tau_l\, \langle N_8^*|\hat{\gamma}_a(\vec b_3)| N \rangle - 
\tau_l\tau_a\, \langle N_8^*|\hat{\gamma}_a(\vec b_2)| N \rangle \\ 
 &-& i\sum_{c} f_{alc} \tau_c \, \langle N_8^*|\hat{\gamma}_a(\vec b_1)| N \rangle \Big\}
\hat \Psi_{G\to q\bar q}(\vec r,\alpha)\,\tilde{\chi}_{\bar q}^{\bar \mu} \,, 
\label{Gqq-amp}
\end{eqnarray} 
where $\tilde{\chi}_{\bar q}^{\bar\mu}=i\sigma_y (\chi_{\bar q}^{\bar\mu})^*$, the impact parameters $\vec b_{1,2,3}$ are defined 
in Eq.~(\ref{imp}), $\chi_{q,\bar q}$ are the two-component spinors normalised as
\begin{eqnarray}
\sum_{\mu,\bar\mu}\tilde{\chi}_{\bar q}^{\bar \mu}
\big({\chi_q^\mu}^\dagger\big)^*= \hat{1} \,, \qquad
\sum_{\mu,\bar\mu}\big({\chi_q^\mu}^\dagger \hat a  \tilde{\chi}_{\bar q}^{\bar \mu} \big)^*\, 
\big({\chi_q^\mu}^\dagger \hat b  \tilde{\chi}_{\bar q}^{\bar \mu}\big)  = {\rm Tr}\big(\hat a^\dagger \hat b\big) \,,
\end{eqnarray}
and the distribution amplitude of the $G\to q\bar q$ splitting $\hat \Phi_{G\to q\bar q}$ reads
\begin{eqnarray} 
\label{Gqq-wf}
\hat \Psi_{G\to q\bar q}(\vec r,\alpha) = \frac{\sqrt{\alpha_s}}{(2\pi)\sqrt{2}} \, 
\Big\{m_q(\vec e\cdot \vec \sigma)+i(1-2\alpha)(\vec\sigma\cdot \vec n)(\vec e\cdot \vec\nabla)-
(\vec e\times \vec n)\cdot \vec\nabla \Big\}\,K_0(\epsilon\, r) \,, \nonumber
\end{eqnarray}
with $\epsilon^2=m_q^2-\alpha \bar{\alpha} m_G^2$. 

When taking square of the total inclusive $G+N\to q\bar q+X$ amplitude
\begin{eqnarray}
\overline{ |A|^2 }(\vec r_1;\vec r_2)\equiv \frac{1}{8}\,\int d^2 s\, d\{X\}\sum_{\lambda_{*},l,\mu,\bar\mu}
\Big\langle A^{\mu\bar\mu}_{l}(\vec s, \vec r_1) \big(A^{\mu\bar\mu}_{l}\big)^\dagger(\vec s, \vec r_2)\Big\rangle 
\label{A2av}
\end{eqnarray}
one performs an averaging over color index and, implicitly, over polarisation $\lambda_*$ 
of the projectile gluon $G$ as well as valence quarks and their relative coordinates 
in the target nucleon. The corresponding inclusive cross section
\begin{eqnarray} \nonumber
&& \frac{d^3\sigma_{\rm incl}(GN\to q\bar qX)}{d(\ln\alpha)d^2\kappa} = \frac{1}{(2\pi)^2}
\int d^2rd^2r'\, e^{i(\vec r - {\vec r}\,')\vec\kappa} \\  && \qquad \times \overline{\sum}\hat \Psi_{G\to q\bar q}(\vec r,\alpha) 
\hat \Psi^\dagger_{G\to q\bar q}({\vec r}\,',\alpha)\Sigma^{G\to q\bar q}_{\rm eff}(\vec r,{\vec r}\,',\alpha)
\label{Gqq-incl}
\end{eqnarray}
where
\begin{eqnarray}
\overline{\sum}\hat \Psi^*_{G\to q\bar q} (\alpha, \vec{r}) \hat \Psi_{G\to q\bar q} (\alpha, \vec{r}\,') & = & 
\frac{\alpha_s}{4\pi^2} \left[m_q^2 K_0(\epsilon\, r) K_0(\epsilon\, r') \right. \nonumber \\
& \, & \left. +  (\alpha^2 + \bar \alpha^2) \epsilon^2 \frac{\vec{r}\cdot \vec{r}\,'}{r r'} 
K_1(\epsilon\, r)K_1(\epsilon\, r')\right] \,,
\label{Gqq-wf}
\end{eqnarray}
and the effective dipole cross section reads
\begin{eqnarray} \nonumber
\Sigma^{G\to q\bar q}_{\rm eff}(\vec r,{\vec r}\,',\alpha) & = & 
\frac{1}{2}\Big\{
\sigma_{Gq\bar q}(-\alpha\vec r,\bar\alpha \vec r\,') + 
\sigma_{Gq\bar q}(\bar\alpha\vec r,-\alpha \vec r\,') \\ 
&-&\sigma_{q\bar q}\big(\alpha(\vec r - \vec r\,')\big) - 
\sigma_{q\bar q}\big(\bar\alpha(\vec r - \vec r\,')\big)   \Big\} \,,
\label{Gqq-sig-eff}
\end{eqnarray}
in terms of the $Gq\bar q$ cross section defined in Eq.~(\ref{GG-Gqq}).

Analogically, the amplitude for inclusive $GN\to G_1G_2 N_8^*$ process in gluon-target scattering reads 
(see Fig.~\ref{fig:dijets-GG/qq} (second row))
\begin{eqnarray}
\nonumber
\hat{A}_{l'ls}(GN\to G_1G_2 N_8^*) &=& \frac{1}{2\sqrt{6}}\,\sum_{a,b} 
\Big\{f_{lab}f_{l'sb}\, \langle N_8^*|\hat{\gamma}_a(\vec b_3)| N \rangle - 
f_{lsb}f_{l'ab}\, \langle N_8^*|\hat{\gamma}_a(\vec b_2)| N \rangle \\ 
 &+& f_{l'lb}f_{asb}\, \langle N_8^*|\hat{\gamma}_a(\vec b_1)| N \rangle \Big\}
\Psi_{G\to G_1G_2}(\vec r,\alpha) \,, 
\label{GGG-amp}
\end{eqnarray} 
where $s,l',l$ are the color indices of the initial $G$ and final $G_1,G_2$ gluons having polarisations $\vec e$, $\vec e_1$, $\vec e_2$, 
respectively, and the $G\to G_1G_2$ distribution amplitude is given by
\begin{eqnarray} \nonumber
\Psi_{G\to G_1G_2}(\vec r,\alpha) &=& \frac{\sqrt{8\alpha_s}}{\pi}\,
\Big\{ \alpha\bar{\alpha} (\vec{e}^{\,*}_1\cdot\vec{e}^{\,*}_2)(\vec{e}\cdot \vec\nabla) - 
\alpha (\vec{e}^{\,*}_1\cdot\vec{e})(\vec{e}^{\,*}_2\cdot \vec\nabla) \\ 
&-& \bar{\alpha} 
(\vec{e}^{\,*}_2\cdot\vec{e})(\vec{e}^{\,*}_1\cdot \vec\nabla) \Big\}\,K_0(\omega\, r)
\label{GGG-wf}
\end{eqnarray}
with $\omega^2=m_G^2(1-\alpha \bar{\alpha})$, such that
\begin{eqnarray} \nonumber
&&\frac{d^3\sigma_{\rm incl}(GN\to G_1G_2X)}{d(\ln\alpha)d^2\kappa} = \frac{1}{(2\pi)^2}
\int d^2rd^2r'\, e^{i(\vec r - {\vec r}\,')\vec\kappa} \\ && \qquad \times \overline{\sum}\Psi_{G\to G_1G_2}(\vec r,\alpha) 
\Psi^\dagger_{G\to G_1G_2}({\vec r}\,',\alpha)\Sigma^{G\to G_1G_2}_{\rm eff}(\vec r,{\vec r}\,',\alpha)
\label{GGG-incl}
\end{eqnarray}
where 
\begin{eqnarray} \nonumber
\overline{\sum}\Psi_{G\to G_1G_2} (\alpha, \vec{r}) \Psi^\dagger_{G\to G_1G_2} (\alpha, \vec{r}\,') & = & 
\frac{8\alpha_s\,\omega^2}{\pi^2}\frac{\vec{r}\cdot \vec{r}\,'}{r\,r'}\,(1-\alpha\bar\alpha)^2 \\
&\times& K_1(\omega\, r)K_1(\omega\, r')\,,
\label{GGG-wf-2}
\end{eqnarray}
and the effective dipole cross section reads
\begin{eqnarray} \nonumber
\Sigma^{G\to G_1G_2}_{\rm eff}(\vec r,{\vec r}\,',\alpha) & = & \frac{9}{16} 
\Big\{\sigma_{q\bar{q}} (\alpha \vec{r}) + \sigma_{q\bar{q}} (\bar\alpha \vec{r}) + 
\sigma_{q\bar{q}} (\alpha \vec{r}\,') + \sigma_{q\bar{q}} (\bar\alpha \vec{r}\,') \\ 
&+& \sigma_{q\bar{q}} (\bar\alpha \vec{r}+\alpha\vec{r}\,') + 
\sigma_{q\bar{q}} (\alpha \vec{r}+\bar\alpha\vec{r}\,') \nonumber \\
&-& 
2\sigma_{q\bar{q}} (\alpha |\vec{r}-\vec{r}\,'|) - 
2\sigma_{q\bar{q}} (\bar\alpha |\vec{r}-\vec{r}\,'|) \Big\}\,.
\label{GGG-sig-eff}
\end{eqnarray}
In the limit of small $\alpha \ll 1$, it can be represented as
\begin{eqnarray}
\Sigma^{G\to G_1G_2}_{\rm eff}(\vec r,{\vec r}\,',\alpha)\Big|_{\alpha\to0} =
\frac{1}{2}\Big\{ \sigma_{3G} (\vec{r},\alpha) + \sigma_{3G} (\vec{r}\,',\alpha) - 
\sigma_{3G} (\vec{r}-\vec{r}\,',\alpha)\Big\} \,,
\end{eqnarray}
in terms of effective 3-gluon cross section
\begin{eqnarray}
\sigma_{3G} (\vec{r},\alpha) = \frac{9}{8} \Big\{ \sigma_{q\bar{q}} (\vec{r}) + \sigma_{q\bar{q}} (\alpha \vec{r}) 
+ \sigma_{q\bar{q}} (\bar\alpha \vec{r}) \Big\} \simeq \sigma_{\rm GG} (\vec{r})\equiv \frac{9}{4} \sigma_{q\bar{q}} (\vec{r})\,, \quad \alpha\ll 1 \,.
\end{eqnarray}
Then for small $\alpha\ll 1$, the ratio between the $qG$ and $GG$ total cross sections
\begin{eqnarray}
\frac{\sigma_{G\to G_1G_2}}{\sigma_{q\to qG}} = 6
\end{eqnarray}
is given by the color factors only.

\section{Hard vs soft dipole scattering}
\label{Sec:dipole-CS}

The phenomenological dipole cross section is the essential ingredient of the color dipole approach \cite{Kopeliovich:1981pz}.
Typically, it is introduced in the form of a saturated ansatz \cite{GBW}
\begin{eqnarray}
\sigma_{q\bar q}(x,\vec r) = \sigma_0\Big[ 1 - e^{-\frac{{r}^2}{R_0^2(x)}} \Big] \,,
\label{ansatz}
\end{eqnarray}
whose Bjorken $x$-dependence is phenomenologically motivated by a wealth of experimental data from HERA. Its parameterisation 
fitted to HERA DIS data known as the Golec-Biernat-W\"usthoff (GBW) model reads
\begin{eqnarray} \nonumber
&& R_0^2\equiv \frac{4}{Q_s^2}\,, \quad Q_s^2(x) \equiv Q_0^2\left( \frac{x_0}{x} \right)^\lambda \,, \quad 
Q_0^2 = 1\,\mathrm{GeV}^2\,, \\
&& x_0 = 4.01 \times 10^{-5}\,, \quad \lambda = 0.277\,, \quad \sigma_0 = 29\, \mathrm{mb} \,.
\label{GBW-params}
\end{eqnarray}
Such a parameterisation, although does not account for the QCD evolution of the target gluon density, still provides a good overall description
of many observables in lepton-hadron and hadron-hadron collisions at small $x\lesssim 0.01$ and at not very large $Q^2$.

During past two decades, various saturation-based parameterisations for the universal dipole cross section that accommodate QCD 
evolution has been proposed based upon the observation of Refs.~\cite{Blaettel:1993rd,Frankfurt:1993it,Frankfurt:1996ri,bartels} 
that the saturation scale is proportional to the collinear gluon density in the target nucleon
\begin{eqnarray}
Q_s^2 = Q_s^2(x,\mu^2) \propto \alpha_s(\mu^2)\,xg(x,\mu^2) \,,
\end{eqnarray}
with the hard scale $\mu^2\sim 1/r^2$. Provided that this scale is not too large, like in the case under consideration of $p_T$-integrated observables 
of dijet production, we will not explicitly incorporate such a dependence, but for the sake of simplicity, will employ the GBW parameterisation 
\cite{GBW}.

Besides saturation, a common property of all the dipole parameterisation is the color transparency limit \cite{Kopeliovich:1981pz},
meaning that a point-like colorless object does not interact with external color fields, i.e.
\begin{eqnarray} 
\sigma_{q\bar q}(x,\vec r) \simeq \sigma_0\,\frac{r^2}{R_0^2(x)} \,, \qquad  r^2 \ll R_0^2(x) \,,
\label{CT}
\end{eqnarray}
which concerns the hard dipole scattering at the scale $\mu\gg Q_s(x)$. The quadratic dependence of 
the universal dipole cross section $\sigma_{q\bar q}\propto r^2$ is a straightforward consequence of 
gauge invariance and non-Abelian nature of QCD.

Integrating the inclusive dijet cross section (\ref{incl}) over $\vec{\kappa}$, we write,
\begin{eqnarray}
\frac{d\sigma_{\rm incl}(qN\to qGX)}{d(\ln\alpha)} = 
\int d^2r\, |\Psi_{q\to qG}(\vec r,\alpha)|^2\, \Sigma^{q\to qG}_{\rm eff}(\vec r,{\vec r},\alpha) \,.
\end{eqnarray}
Here the effective dipole cross section in the small dipole size limit $r\ll R_0(x_2)$
\begin{eqnarray}
\Sigma^{q\to qG}_{\rm eff}(\vec r,{\vec r},\alpha) \simeq  {\cal K}^{q\to qG}_{\rm incl}(x_2,\alpha)\,r^2 \,, \quad 
{\cal K}^{q\to qG}_{\rm incl}(x_2,\alpha)=\frac{\sigma_0}{R_0^2(x_2)}\cdot 
\Big[\frac{9}{4}\bar{\alpha} +\alpha^2\Big]\,, \quad x_2=\frac{M^2}{x_q\,s}\,,
\label{x2}
\end{eqnarray}
and $s$ is the nucleon-nucleon c.m. energy squared. The fully differential cross section for the inclusive 
$q+G$ production in this approximation takes a very simple form
\begin{eqnarray}
\frac{d\sigma^{\rm NN}_{\rm incl}}{d\Omega}\simeq \frac{{\cal K}^{q\to qG}_{\rm incl}(x_2,\alpha)}{(2\pi)^2}\,q(x_q,\mu^2)\,
\int d^2rd^2r'\, e^{i\vec \kappa (\vec{r}-\vec{r}\,')}\,(\vec{r}\cdot \vec{r}\,')\,\Psi_{q\to qG}(\vec r,\alpha)\Psi_{q\to qG}^\dagger(\vec r\,',\alpha) \,,
\label{inclCS}
\end{eqnarray}
where the phase space volume element is 
\begin{eqnarray}
d\Omega=dx_q\,d\ln\alpha\,d^2\kappa \,.
\label{PS-element}
\end{eqnarray}
For the gluon-initiated processes $G\to q\bar q$ 
and $G\to G_1G_2$ we have
\begin{eqnarray}
 {\cal K}^{G\to q\bar q}_{\rm incl}(x_2,\alpha)=\frac{\sigma_0}{R_0^2(x_2)}\cdot 
\Big[1-\frac{9}{4}\alpha\bar{\alpha}\Big] \,, \qquad  
{\cal K}^{G\to G_1G_2}_{\rm incl}(x_2,\alpha)=\frac{9\sigma_0}{4R_0^2(x_2)}\cdot 
\Big[1-\alpha\bar{\alpha}\Big] \,,
\label{glu-in}
\end{eqnarray}
respectively.

In the soft limit $Q^2\to\Lambda_{\rm QCD}^2$ one can reach very small values of $x$ defined in Eq.~(\ref{x2}) even at low energies. 
This signals about inappropriate use of variable $x_2$ in this limit.
In soft and semi-soft reactions such as pion-proton scattering, or diffractive processes Drell-Yan and gluon radiation, the saturation scale 
depends on the gluon-target collision c.m. energy squared $\hat{s}=x_q\,s$ which is a more appropriate variable than the Bjorken $x$.
Such reactions are characterised by the associated scale $Q^2\sim \Lambda_{\rm QCD}^2 \sim 1/R_{\rm had}^2$ at the soft 
hadronic scale $R_{\rm had}$. Keeping the saturated ansatz of the dipole cross section (\ref{ansatz}), the corresponding
parameterisation for $\sigma_0 \to \overline{\sigma}_0(\hat{s})$ and $R_0 \to \overline{R}_0(\hat{s})$ has been found 
in Ref.~\cite{Kopeliovich:1999am}
\begin{eqnarray}\nonumber
 \overline{R}_0(\hat s)=0.88\,\mathrm{fm}\,(s_0/\hat s)^{0.14}\,,\quad
 \overline{\sigma}_0(\hat s)=\sigma_{\rm tot}^{\pi p}(\hat s)
 \Big(1+\frac{3\overline{R}_0^2(\hat s)}{8\langle r_{\rm ch}^2 \rangle_{\pi}}\Big)\,.
 \label{KST-params}
\end{eqnarray}
in terms of the pion-proton total cross section given by $\sigma_{\rm tot}^{\pi p}(\hat s)=23.6(\hat s/s_0)^{0.08}$ mb \cite{barnett},
$s_0=1000\,{\rm GeV}^2$, the mean pion charge radius squared $\langle r_{\rm ch}^2 \rangle_{\pi}=0.44$ fm$^2$ \cite{amendolia}. 
This parameterisation describes well the HERA data for the proton structure function at medium-high scales up to $Q^2\sim 10$ GeV$^2$. 
The model (\ref{KST-params}) will be referred below to as the KST model and used in our analysis of diffractive dijet production in 
high-energy hadronic collisions.

\section{Single-diffractive dijets production}
\label{Sec:jj-SD}

The main contribution to the diffractive dijets production cross section at very forward rapidities is given by 
the diffractive gluon bremsstrahlung off the projectile valence quarks $q\to qG$ as is demonstrated  in Fig.~\ref{fig:quark-exc} 
(for an analogous discussion in the case of diffractive Abelian bremsstrahlung, see Refs.~\cite{Kopeliovich:2006tk,Pasechnik:2011nw,
Pasechnik:2012ac,Pasechnik:2014lga,Pasechnik:2015fxa}). At hadron colliders such as Tevatron, however, the jet rapidities 
may extend down to central values where the contribution from diffractive gluon excitation, given by the gluon splitting 
subprocesses $G\to q\bar q$ and $G\to GG$, become important. 
Diffractive excitation of the projectile sea-quarks also contributes, but negligibly less compared with gluons.
In what follows, we discuss all these reactions on the
same footing and derive the corresponding SD cross sections.

\subsection{Diffractive excitation of a projectile quark}

\begin{figure*}[!h]
 \centerline{\includegraphics[width=0.85\textwidth]{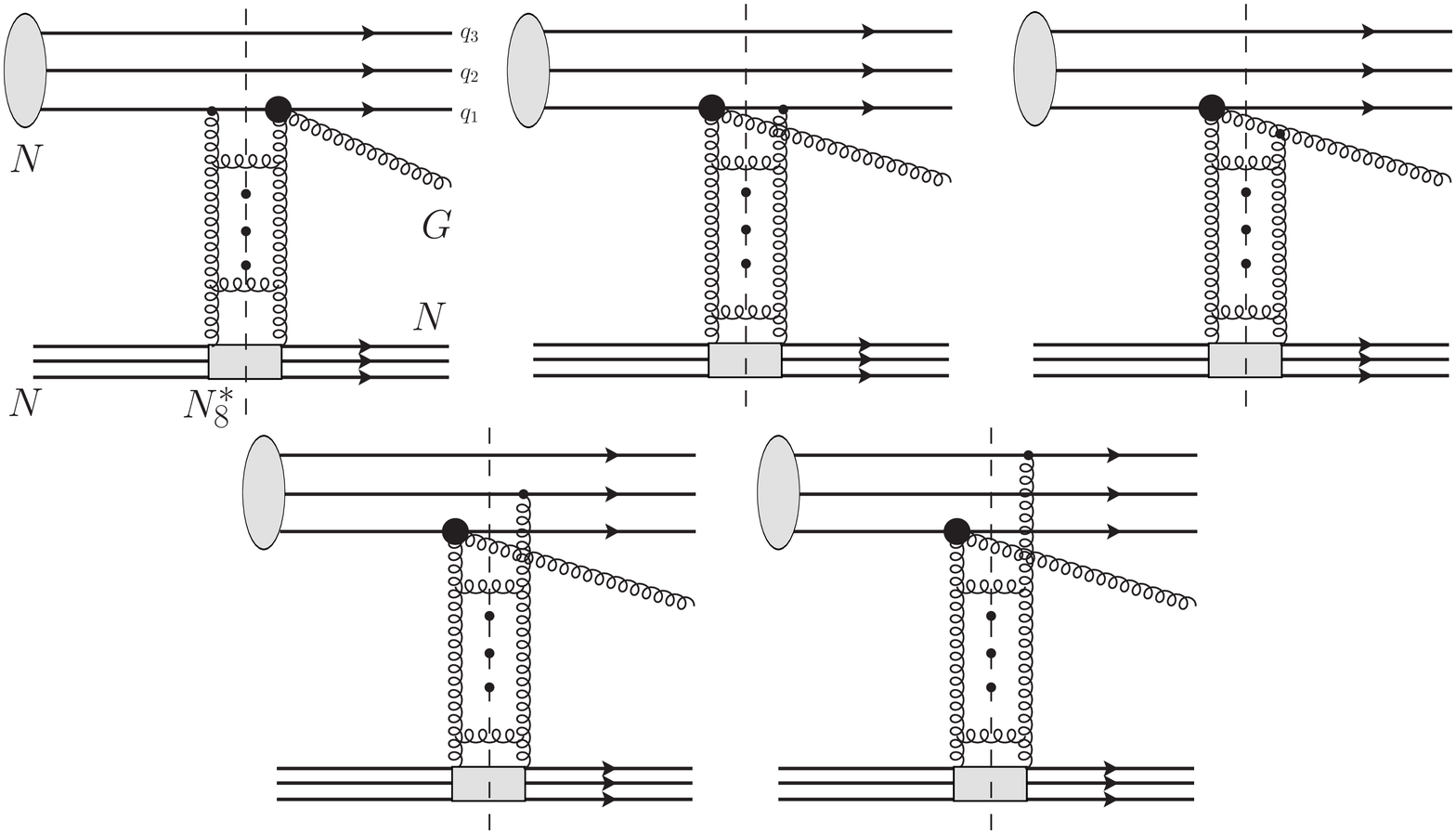}}
   \caption{
   \small Dijet production from diffractive quark excitation in $NN$ collisions. Additional graphs 
   come from $q_1 \leftrightarrow q_2$ and $q_1 \leftrightarrow q_3$ permutations. Large filled circle 
   corresponds to three perturbative leading-order contributions depicted in Fig.~\ref{fig:hard-dijets}.
   }
 \label{fig:quark-exc}
\end{figure*}

The hadron-level SD amplitude with the gluon bremsstrahlung process $q\to qG$ can be conveniently decomposed 
into three parts,
\begin{eqnarray}
\hat{A}_{l}^{q_i,{\rm SD}} = \hat{A}^{q_i,{\rm (I)}}_l + \hat{A}^{q_i,{\rm (II)}}_l + \hat{A}^{q_i,{\rm (III)}}_l \,, \qquad 
\hat{A}_{l}^{{\rm SD}}\equiv \sum_{i=1}^3\hat{A}_{l}^{q_i,{\rm SD}} \,,
\label{q1-SD}
\end{eqnarray}
with color index $l$ of the gluon by the projectile quark $q_i$ ($i=1,2,3$). In what follows, we keep the earlier introduced notation notation for the total  transverse 
momentum transfer $\vec{q}_\perp$ conjugated to the impact parameter $\vec{b}$, like in the inclusive case.

First, let us consider the three graphs in the upper row of Fig.~\ref{fig:quark-exc} corresponding to the amplitude $\hat{A}^{q,\rm (I)}_l$ 
of  diffractive gluon radiation in nucleon-nucleon scattering $NN\to (3q)_8GN$. One of the $t$-channel gluons, which couples to the hard 
scale $\mu^2$ (large filled circle in Fig.~\ref{fig:quark-exc}), we call ``active'' gluon. In order to keep the whole $t$-channel exchange 
colorless, as is required in a diffractive process, there should be an additional effective color 
octet exchange between any projectile or produced parton and the target, which we call ``screening'' gluon. Then, the amplitude 
$\hat{A}^{q,\rm (I)}_l$ is related to the amplitude of diffractive gluon radiation in the quark-nucleon $qN\to qGN$ scattering $\hat{A}^{q}_{l}$ 
as follows,
\begin{eqnarray}
\hat{A}^{q,\rm (I)}_l = -2i\,\langle (3q)_8 |\hat{A}^{q}_{l}| (3q)_1 \rangle\,\Phi_{N\to (3q)_1} \,,
\end{eqnarray}
where $\Phi_{N\to (3q)_1}(\vec r_1,\vec r_2,\vec r_3;\alpha_1,\alpha_2,\alpha_3)$ is the nucleon wave function 
describing a fluctuation of the projectile nucleon into a colorless system $(3q)_1$ of three valence quarks $i=1,2,3$ with transverse 
positions and light-cone momentum fractions $\{\vec r_i\}$ and $\{\alpha_i\}$ respectively.

At high energies, the diffractive amplitude is predominantly imaginary. So, using the generalised optical theorem for the unitarity cut 
(dashed vertical lines in Fig.~\ref{fig:quark-exc}) between 
the ``active'' and ``screening'' gluons, and summing up the corresponding contributions, 
we obtain
\begin{eqnarray} 
\nonumber
\hat{A}^{q}_{l}&=&\frac{i}{2} \sum_{N_8^*} \Big[ \hat{A}_l^\dagger(qGN\to qN_8^*)\hat{A}(qN\to qN_8^*) \\
&+& \sum_{l'}\hat{A}_{ll'}^\dagger(qGN\to qG'N_8^*)\hat{A}_{l'}(qN\to qG'N_8^*) \Big] \,.
\label{qN}
\end{eqnarray}
Here the first term corresponds to the first diagram in Fig.~\ref{fig:quark-exc}, and the second one to the sum of the
second and third diagrams, with explicit summation over intermediate color index of the $G'$ gluon $l'$ and nucleon 
octet-charged remnant $N_8^*\equiv (3q)_8$.

In the impact parameter representation, the scattering amplitude $\hat{A}_{l'}(qN\to qG'N_8^*)$ has the same form as the inclusive 
production amplitude Eq.~(\ref{qG-incl-A}), while the other amplitudes read
\begin{eqnarray} \nonumber
\hat{A}(qN\to qN_8^*)&=&\sum_{a} \tau_a\langle N_8^*|\hat{\gamma}_a(\vec b_1)| N \rangle \,, \\
\hat{A}_{ll'}^{\dagger}(qGN\to qG'N_8^*)&=&\delta_{ll'}\sum_{a} \tau_a\langle N_8^*|\hat{\gamma}_a(\vec b_2)| N \rangle - 
\sum_{a} if_{ll'a}\langle N_8^*|\hat{\gamma}_a(\vec b_3)| N \rangle \,,  \nonumber \\
\hat{A}^\dagger_{l}(qGN\to qN_8^*)&=&\frac{\sqrt{3}}{2}\sum_{a}\Big[\tau_l\tau_a\langle N_8^*|\hat{\gamma}_a(\vec b_1)| N \rangle - 
\tau_a\tau_l\langle N_8^*|\hat{\gamma}_a(\vec b_2)| N \rangle \nonumber \\ 
&-&\sum_{b} if_{lab}\tau_b\langle N_8^*|\hat{\gamma}_a(\vec b_3)| N \rangle\Big]\, 
\Psi_{q\to qG}(\vec r,\alpha) \,.
\label{other-amps}
\end{eqnarray}
Then, using Eq.~(\ref{averaging}) one arrives at the SD amplitude of $qN\to qGN$ process
\begin{eqnarray}
\hat{A}^{q}_{l}=\frac{3i\sqrt{3}}{16}\,\tau_{l}\,\Psi_{q\to qG}(\vec r,\alpha)
\Big\{ \frac43\Big(\phi({\vec b}_1,{\vec b}_1)-\phi({\vec b}_2,{\vec b}_2)\Big) + 
3\Big(\phi({\vec b}_2,{\vec b}_3)-\phi({\vec b}_3,{\vec b}_3)\Big)\Big\} \,,
\end{eqnarray}
which is infrared finite and vanishes in the color transparecy limit $\vec r\to 0$, despite the divergency in the  
amplitude $\phi(\vec b_k,\vec b_l)$. The symmetry properties of $\phi(\vec b_k,\vec b_l)$ in particular imply,
\begin{eqnarray*}
\int d^2 b \sum_{i} {\cal C}_i \phi(\vec d_i,\vec d_i)=0 \qquad \mathrm{for} \qquad \sum_{i} {\cal C}_i = 0 \,, \quad \vec d_i = \vec b + \vec y_i
\end{eqnarray*}
for any $\vec y_i$ such that in the forward diffractive scattering limit $\vec{q}_\perp\to0$ we finally have,
\begin{eqnarray}
\hat{A}^{q}_{l}({\vec q}_\perp,\vec\kappa)\Big|_{q_\perp\to 0} = \int d^2b\, d^2r\, e^{i\vec\kappa \vec r}\,
\hat{A}^{q}_{l}(\vec b,\vec r) = -\frac{9i\sqrt{3}}{32}\,\tau_{l}\,\int d^2r e^{i\vec\kappa \vec r}\,\Psi_{q\to qG}(\vec r,\alpha)
\sigma_{q\bar q}(\vec r) \,.
\end{eqnarray}

For small dipoles $r^2 \ll R_0^2(x)$, the diffractive amplitude transforms to
\begin{eqnarray}
\hat{A}^{q}_{l}({\vec q}_\perp,\vec\kappa)\Big|_{q_\perp\to 0} 
\simeq \frac{9i\sqrt{3}}{32}\,\frac{\sigma_0}{R_0^2(x)}\,\tau_{l}\, ( \vec{\nabla}_{\kappa} 
\cdot \vec{\nabla}_{\kappa} ) \, \hat{\tilde{\Psi}}_{q\to qG}(\vec \kappa,\alpha) \,,
\end{eqnarray}
where $\vec{\nabla}_{\kappa} = \partial/\partial \vec{\kappa}$, such that
\begin{eqnarray}
( \vec{\nabla}_{\kappa} \cdot \vec{\nabla}_{\kappa} ) \, \hat{\tilde{\Psi}}_{q\to qG}(\vec \kappa,\alpha) 
\simeq \frac{8\sqrt{\alpha_s}}{\sqrt{3}}\chi_f^\dagger \Big\{i\alpha \vec{e}_G\cdot[\vec{\sigma}\times \vec{\kappa}] - 
(2-\alpha) (\vec{e}_{G}\cdot\vec{\kappa}) \Big\}\chi_i\, \frac{1}{\kappa^4}\,.
\end{eqnarray}

Diffractive quark-to-dijet excitation offers another test of factorization. Indeed, in this case there are no spectator partons, which would cause 
a suppressive gap survival probability, which is usually identified as the reason for factorisation breaking. However, factorisation fails even 
without such a gap survival factor. Indeed, the corresponding differential cross section of the SD dijets production in the quark-nucleon scattering 
$qN\to qGN$ has the following form (c.f.~Ref.~\cite{Kopeliovich:1999am})
\begin{eqnarray} \nonumber
&&\frac{d^3\sigma_{\rm SD}(qN\to qGN)}{d(\ln\alpha)d^2q_\perp}\Big|_{q_\perp\to 0} =
\frac13\,\frac{1}{(2\pi)^2} \int \frac{d^2\kappa}{(2\pi)^2}\sum_{l}{\rm Tr}\,\Big[\hat{A}^{q}_{l}({\vec q}_\perp,\vec\kappa)
\hat{A}^{q\dagger}_{l}({\vec q}_\perp,\vec\kappa)\Big]\Big|_{q_\perp\to 0} \\
&& \qquad\qquad\quad = \frac{1}{16\pi^2}\int d^2r \Big| \Psi_{q\to qG}(\vec r,\alpha)\tilde{\sigma}_{q\bar q}(\vec r) \Big|^2 \,, \quad
\tilde{\sigma}_{q\bar q}(\vec r)\equiv \frac{9}{8}\sigma_{q\bar q}(\vec r)\,,
\label{SD-CS-quark}
\end{eqnarray}
where the factor $1/3$ stands for averaging over colors of the projectile quark. In the color transparency (or large radiated gluon transverse 
momentum) limit, we get
\begin{eqnarray} \nonumber
&&\frac{d^3\sigma_{\rm SD}(qN\to qGN)}{d(\ln\alpha)d^2\kappa\,d^2q_\perp}\Big|_{q_\perp\to 0} = 
\frac{1}{16\pi^2}\, \frac{81}{64}\, \frac{\sigma_0^2}{R_0^4(x)}\, \frac{1}{(2\pi)^2}\, \frac{1}{2}\sum_{i,f,\lambda_G}
\Big| ( \vec{\nabla}_{\kappa} \cdot \vec{\nabla}_{\kappa} ) \, \hat{\tilde{\Psi}}_{q\to qG}(\vec \kappa,\alpha) \Big|^2 \,,
\end{eqnarray}
where the amplitude squared (averaged over the incoming quark helicities) reads explicitly
\begin{eqnarray}
\frac{1}{2}\sum_{i,f,\lambda_G}
\Big| ( \vec{\nabla}_{\kappa} \cdot \vec{\nabla}_{\kappa} ) \, \hat{\tilde{\Psi}}_{q\to qG}(\vec \kappa,\alpha) \Big|^2 = 
\frac{128\alpha_s}{3}\, \frac{2-\alpha(2-\alpha)}{\kappa^6} \,.
\label{kappa-dep}
\end{eqnarray}
If factorisation were true, the diffractive structure functions are nearly scale independent (only logarithmically). Therefore, all the dependence 
on $\kappa$ comes from the hard parton-parton scattering, i.e. should scale as $1/\kappa^4$, in apparent contradiction with the result (\ref{kappa-dep}).

Coming to the hadron-level SD amplitude $NN\to (3q)_8GN$, we define the impact parameters for a gluon radiation off the $i$th projectile 
quark ($i=1,2,3$) in terms of its transverse position $\vec r_i$ relative to the impact parameter $\vec b$ as follows,
\begin{eqnarray}
\vec{b}^{(i)}_1\equiv \vec{b}+\vec r_i\,, \qquad \vec{b}^{(i)}_2\equiv \vec{b} + \vec r_i - \alpha\vec \rho_i \,, \qquad 
\vec{b}^{(i)}_3\equiv \vec{b}+\vec r_i + \bar\alpha\vec \rho_i \,, \qquad \vec \rho_i = \vec \rho - \vec r_i \,,
\end{eqnarray}
where the difference between transverse coordinates of the radiated gluon, $\vec \rho$, and the position of the parent projectile quark is $\vec r_i$.
Thus, the first term in Eq.~(\ref{q1-SD}) for gluon radiation off the projectile quark $q_1$ can be presented as,
\begin{eqnarray}
\hat{A}^{q_1,{\rm (I)}}_l &=& 2i\cdot\frac{i\sqrt{3}}{4}\,\langle (3q)_8 | \tau^{(q_1)}_{l} | (3q)_1 \rangle\,\Phi_{N\to (3q)_1}\, 
\Psi_{q\to qG}(\vec \rho_1,\alpha)\,\tilde{\sigma}_{q\bar q}(\vec \rho_1) \nonumber \\
&=& \frac{if_{lab}}{\sqrt{3}}\,\langle (3q)_8 | \tau^{(q_1)}_{a}\tau^{(q_1)}_{b} | (3q)_1 \rangle\,\Phi_{N\to (3q)_1}\, 
\Psi_{q\to qG}(\vec \rho_1,\alpha)\,\tilde{\sigma}_{q\bar q}(\vec \rho_1)\,,
\end{eqnarray}
Notice that the mean transverse size of the perturbative fluctuation $q\to qG$ with a high-$p_T$ gluon,
controlled by the light-cone distribution function $\Psi_{q\to qG}(\vec \rho_i)$ (see Eq.~(\ref{psi-qG})), is  much smaller than the inter-quark 
separation in the nucleon, which is $R_N\sim 1$ fm, i.e.
\begin{eqnarray}
|\vec \rho_i| \ll |\vec r_{ij}|\sim R_N \,, \qquad i \not = j \,, \qquad \vec r_{ij}\equiv \vec r_i - \vec r_j \,.
\end{eqnarray}

For the second and third terms, $\hat{A}^{q_1,{\rm (II)}}_l$ and $\hat{A}^{q_1,{\rm (III)}}_l$, in Eq.~(\ref{q1-SD}) corresponding to the first and second 
diagrams in the second row of Fig.~\ref{fig:quark-exc}, respectively, we write,
\begin{eqnarray} \nonumber
\hat{A}^{q_1,{\rm (II)}}_l &=& \langle (3q)_8 |\hat{A}^\dagger(q_2N\to q_2N_8^*) \hat{A}_{l}(q_1N\to q_1GN_8^*)  | (3q)_1 \rangle\,\Phi_{N\to (3q)_1} \\
&=& \frac{3\sqrt{3}\Phi_{N\to (3q)_1}\,\Psi_{q\to qG}(\vec \rho_1,\alpha)}{16}\,\Big\{ \langle (3q)_8 | \tau^{(q_2)}_{a}\tau^{(q_1)}_{l}\tau^{(q_1)}_{a} | (3q)_1 \rangle 
\big(\sigma_{q\bar q}(\vec r_{12}) - \sigma_{q\bar q}(\vec r_{12}+\bar\alpha \vec \rho_1) \big) \nonumber \\
&+&
\langle (3q)_8 | \tau^{(q_2)}_{a}\tau^{(q_1)}_{a}\tau^{(q_1)}_{l} | (3q)_1 \rangle 
\big(\sigma_{q\bar q}(\vec r_{12}+\bar\alpha \vec \rho_1) - \sigma_{q\bar q}(\vec r_{12}-\alpha \vec \rho_1)\big) \Big\} \,, \\
\hat{A}^{q_1,{\rm (III)}}_l &=& \langle (3q)_8 |\hat{A}^\dagger(q_3N\to q_3N_8^*) \hat{A}_{l}(q_1N\to q_1GN_8^*)  | (3q)_1 \rangle\,\Phi_{N\to (3q)_1} 
\nonumber \\
&=& \frac{3\sqrt{3}\Phi_{N\to (3q)_1}\,\Psi_{q\to qG}(\vec \rho_1,\alpha)}{16}\,\Big\{ \langle (3q)_8 | \tau^{(q_3)}_{a}\tau^{(q_1)}_{l}\tau^{(q_1)}_{a} | (3q)_1 \rangle 
\big(\sigma_{q\bar q}(\vec r_{13}) - \sigma_{q\bar q}(\vec r_{13}+\bar\alpha \vec \rho_1) \big) \nonumber \\
&+&
\langle (3q)_8 | \tau^{(q_3)}_{a}\tau^{(q_1)}_{a}\tau^{(q_1)}_{l} | (3q)_1 \rangle 
\big(\sigma_{q\bar q}(\vec r_{13}+\bar\alpha \vec \rho_1) - \sigma_{q\bar q}(\vec r_{13}-\alpha \vec \rho_1)\big) \Big\} \,,
\end{eqnarray}
in terms of the partial amplitudes given in Eqs.~(\ref{qG-incl-A}) and (\ref{other-amps}).

In practical calculations, it is convenient to employ the following relation
\begin{eqnarray}
(\tau^{(q_1)}_a+\tau^{(q_2)}_a+\tau^{(q_3)}_a)  | (3q)_1 \rangle = 0
\end{eqnarray}
and a more generic formula for cyclic permutations $\{q_1,\,q_2,\,q_3\}$ of the products of $\tau^{(q_j)}$-matrices,
\begin{eqnarray}
\Big( P^{q_1}P^{q_2}P^{q_3} + P^{q_2}P^{q_3}P^{q_1} + P^{q_3}P^{q_1}P^{q_2} \Big) | (3q)_1 \rangle = 0 \,, \qquad 
P^{q_j} = \tau_a^{q_j}\tau_b^{q_j}\dots \,,
\end{eqnarray}
for any product of $\tau$-matrices $P^{q_j}$ along a quark line $q_j$, $j=1,2,3$ or unity. Averaging over the nucleon state 
$| (3q)_1 \rangle$ in the SD amplitude squared is performed as follows
\begin{eqnarray*}
\langle (3q)_1 | A(\tau^{(q_1)})B(\tau^{(q_2)})C(\tau^{(q_3)}) | (3q)_1 \rangle &=& \frac16 \Big( \rm{Tr}[A]\rm{Tr}[B]\rm{Tr}[C] + 
\rm{Tr}[ABC] + \rm{Tr}[ACB] \\
&-& \rm{Tr}[A]\rm{Tr}[BC] - \rm{Tr}[B]\rm{Tr}[AC] - \rm{Tr}[C]\rm{Tr}[AB]  \Big)
\end{eqnarray*}
where $A,B,C$ are any products of $\tau$-matrices corresponding to $q_{1,2,3}$ projectile quarks, respectively.

Assuming the saturated form of the dipole cross section, up to the terms containing the first power of $\rho_i \ll r_{ij}$, we write for two distinct cases
\begin{eqnarray}
\mathrm{hard}\;\mathrm{regime}:
&&\sigma_{q\bar q}(\vec \rho_i)\simeq \sigma_0\,\frac{\rho_i^2}{R_0^2(x)} \,, \\ 
\mathrm{soft}\;\mathrm{regime}:
&&\sigma_{q\bar q}(\vec r_{ij})-
\sigma_{q\bar q}(\vec r_{ij} - \alpha \vec \rho_i) \simeq  2\alpha (\vec \rho_i \cdot \vec r_{ij})\,
\frac{\overline{\sigma}_0(\hat{s})}{\overline{R}_0^2(\hat{s})}\,e^{-r_{ij}^2/\overline{R}_0^2(\hat{s})} \,,
\end{eqnarray}
where the sets of parameters in the universal dipole cross section $\{\sigma_0,\,R_0(x)\}$ and $\{\overline{\sigma}_0(\hat{s}),\,
\overline{R}_0(\hat{s})\}$ are determined in the hard-dipole scattering (GBW model (\ref{GBW-params})) and soft-dipole scattering (KST model 
(\ref{KST-params})) regimes, respectively. Provided that $\rho_i \ll r_{ij}$, we can safely neglect the interference terms for gluon 
emissions off different projectile quarks, such that only the diagonal product,
\begin{eqnarray}
|\Psi_{q\to qG}(\vec \rho_i,\alpha)|^2&=&\frac{4}{3}\frac{\alpha_s(\mu^2)}{2\pi^2}\Big\{m_q^2\alpha^4\,K_0^2(\tau\rho_i)
+\Big[1+(1-\alpha)^2\Big]\tau^2\,K_1^2(\tau\rho_i)
\Big\},
\end{eqnarray}
contributes to the final result for the (integrated) SD cross section.

When computing the SD amplitude squared we have to use the completeness relation $| N^*_8 \rangle \langle N^*_8 | = 1$ 
which accounts for the momentum conservation for the nucleon remnant wave function $\Psi_{N^*_8}$. More explicitly,
\begin{eqnarray}\nonumber
&&\sum_{N^*_8}\Psi_{N^*_8}(\vec{r}_1,\vec{r}_2,\vec{r}_3;\{x_q^{1,2,...}\},\{x_g^{1,2,...}\})
\Psi^*_{N^*_8}(\vec{r}\,'_1,\vec{r}\,'_2,\vec{r}\,'_3;\{{x'}_q^{1,2,...}\},\{{x'}_g^{1,2,...}\})\\
&&\phantom{.......}=\,
\delta\bigl(\vec{r}_1-\vec{r}\,'_1\bigr)\delta(\vec{r}_2-\vec{r}\,'_2)
\delta(\vec{r}_3-\vec{r}\,'_3)\prod_{j}\delta(x_{q/g}^j-{x'}_{q/g}^j) \,.
\end{eqnarray}
The wave function of the initial nucleon state $\Phi_{N\to (3q)_1}$ depends on transverse coordinates and fractional momenta 
of all the projectile (valence and sea) quarks and gluons. We assume that all sea quarks and gluons are
 localized within gluonic ``spots'', around the constituent valence quarks, whose small transverse size $\sim 0.3$ fm. 
 The smallness of the spots allows to explain the observed weakness of diffractive gluon radiation \cite{Kopeliovich:1999am} 
 (the puzzling smallness of the triple-Pomeron coupling). There are many other observables confirming such a conclusion \cite{Kopeliovich:2007pq}). 
 This picture supports the popular two-step model \cite{Gluck:1989ze,Gluck:1994uf,Gluck:1998xa}, in which the initial valence-quark 
 distribution function is fixed at a low scale, and then is developed to a higher scale perturbatively by radiative generation of the sea and gluons.
 
Thus, the valence-like spatial wave function of the proton introduced at a low scale, is not subject to further variations as function of scale. 
For the impact parameters $r_1,\,r_2,\,r_3$ of the valence quarks we use the symmetric (normalised) Gaussian parameterisation of 
the valence part of the proton wave function reads
\begin{eqnarray}\nonumber
|\Phi_{N\to (3q)_1}|^2 &=& \frac{3a^2}{\pi^2}e^{-a(r_1^2+r_2^2+r_3^2)}\;{\cal R}\big(\{x_q\},\{x_g\}\big)\\
&\times& \delta(\vec{r}_1+\vec{r}_2+\vec{r}_3)\delta\Big(1-\sum_{j=1}^\infty x_{q}^j-\sum_{j=1}^\infty x_{g}^j\Big) \,, 
\label{psi}
\end{eqnarray}
where $a\equiv \langle r_{\rm ch}^2 \rangle_p^{-1}$ is the inverse proton mean charge radius squared, ${\cal R}$ is 
the generalised parton distribution function in the projectile nucleon. 
In fact, the spatial distribution of the valence quarks in the proton, even the string configuration (triangle vs star shapes), 
are still under debate. Different models were tested in Ref.~\cite{Kopeliovich:2005us} on data of soft diffraction. 
Only the Model IV with symmetric dependence on $r_1,\,r_2,\,r_3$, and the saturated dipole cross section, was found 
to be able to explain the observed puzzling smallness (only few percent of elastic) of the low-mass diffraction cross section. 
The latter is described by the  $\mathbb{P}\mathbb{P}\mathbb{R}$ term in the triple-Regge phenomenology \cite{Kazarinov:1975kw}. 
It corresponds to diffractive excitation of the valence quark skeleton (in contrast to diffractive gluon radiation, giving the 
$\mathbb{P}\mathbb{P}\mathbb{P}$ term), this is why small-mass diffraction is so sensitive to the valence quark distribution.

In the case of diffractive quark excitation we obtain
\begin{eqnarray}
\int \prod_{j\not=1} dx_q^j \prod_{k}dx_g^k\,\delta\Big(1-\sum_{j=1}^\infty x_{q}^j-\sum_{j=1}^\infty x_{g}^j\Big)\;
{\cal R}\big(\{x_q\},\{x_g\}\big) = q(x_q,\mu^2)\,, 
\label{qua}
\end{eqnarray}
in terms of the quark PDF $q(x_q,\mu^2)$ where the projectile (valence or sea) quark momentum fraction is $x_q^1\equiv x_q$.

The SD quark-gluon dijet production cross section in nucleon-nucleon collisions $N+N\to qGX+N$ is found as
\begin{eqnarray}
\frac{d^3\sigma_{\rm SD}}{d\ln \alpha d^2 q_\perp}\Big|_{q_\perp \to 0}=\frac{1}{(4\pi)^2}\,
\int d^2r_1 d^2r_2 d^2r_3\,\prod_{i,j} dx_q^i dx_g^j
\int d^2 \rho\,
\overline{\sum}\hat{A}_{l}^{{\rm SD}}(\hat{A}_{l}^{{\rm SD}})^\dagger \,, 
\label{FSD}
\end{eqnarray}
where $q_\perp^2=-t$. The momentum conservation reduces the integral over the incoming nucleon 
wave function as
\begin{eqnarray}
\int d^2r_1 d^2r_2 d^2r_3\,e^{-a(r_1^2+r_2^2+r_3^2)}\,\delta(\vec{r}_1+\vec{r}_2+\vec{r}_3)=
\frac19\int d^2r_{12}d^2r_{13}\,e^{-\frac{2a}{3}(r_{12}^2+r_{13}^2+\vec{r}_{12}\cdot\vec{r}_{13})}
\end{eqnarray}
such that the basic integrals appearing in the SD cross section
\begin{eqnarray} 
&&\frac{3a^2}{\pi^2}\frac19\int d^2R_1d^2R_2\,e^{-\frac{2a}{3}(R_1^2+R_2^2+\vec{R}_1\cdot\vec{R}_2)}
e^{-(R_i^2+R_j^2)/\overline{R}_0^2}\, (\vec R_{i}\cdot\vec R_{j}) \,, \qquad i,j=1,2,3 \,,
\end{eqnarray}
where $\vec{R}_1\equiv \vec{r}_{12}$, $\vec{R}_2\equiv \vec{r}_{13}$, $\vec{R}_3\equiv \vec{r}_{23}=\vec{r}_{13}-\vec{r}_{12}$, 
can be taken fully analytically. Finally, as usual the SD cross section is the forward limit is inversely proportional 
to the standard Regge-parameterised diffractive $t$-slope, $B_{\rm SD}(s)$, namely,
\begin{eqnarray} \nonumber
&& \frac{d^2\sigma_{\rm SD}}{d\Omega} \simeq \frac{1}{B_{\rm SD}(s)}\,
\frac{d^3\sigma_{\rm SD}}{d\Omega\,dt}\Big|_{t\to0} \,, \\
&& B_{\rm SD}(s)\simeq \langle r_{\rm ch}^2 \rangle_p/3+
2\alpha'_{\Pom}\ln(s/s_1) \,,\quad s_1=1\,{\rm GeV}^2 \,,
\label{slope}
\end{eqnarray}
where $\alpha'_{\Pom}=0.25\,\GeV^{-2}$ and the phase space volume element $d\Omega$ is defined in Eq.~(\ref{PS-element}). 

Following the above footsteps, straightforward calculations lead to the following representation of the fully 
differential cross section for the SD $qG$ production in nucleon-nucleon collisions
\begin{eqnarray} 
\nonumber
\frac{d\sigma^{q\to qG}_{\rm SD}}{d\Omega} &\simeq& \frac{{\cal K}^{q\to qG}_{\rm SD}(s,\hat{s},\alpha)}{(2\pi)^2}\,q(x_q,\mu^2)\,
\int d^2\rho d^2\rho'\, e^{i\vec \kappa (\vec \rho-\vec \rho\,')}\,(\vec{\rho}\cdot \vec{\rho}\,') \\
&\times& \overline{\sum}\hat{\Psi}_{q\to qG}(\vec \rho,\alpha)\hat{\Psi}^{\dagger}_{q\to qG}(\vec \rho\,',\alpha) \,, \label{SD-CS-fin}
\end{eqnarray}
where
\begin{eqnarray} 
{\cal K}^{q\to qG}_{\rm SD}&=&\frac{1}{B_{\rm SD}}\,\frac{9a\overline{\sigma}_0(\hat{s})^2}{256\pi}
\Big\{ {\cal W}_1(\hat{s})\,\Big[ 1 - \frac{2\alpha}{3} +\frac{7\alpha^2}{27} \Big] + 
{\cal W}_2(\hat{s})\,\Big[ 1 + \frac{2\alpha}{3} -\frac{13\alpha^2}{27} \Big] \Big\} \,,
\label{KqqG}
\end{eqnarray}
where the $\hat{s}$-dependent functions read
\begin{eqnarray} 
\nonumber
{\cal W}_1(\hat{s})&=&\frac{8}{(4+a\overline{R}_0^2)^2}+\frac{12}{(12+a\overline{R}_0^2)^2} \,, \qquad \hat{s}=x_q\,s \,, 
\qquad \overline{R}_0=\overline{R}_0(\hat{s}) \,, \\
{\cal W}_2(\hat{s})&=&\frac{6a^2\overline{R}_0^4}{(3+8a\overline{R}_0^2+a^2\overline{R}_0^4)^2} -
\frac{a^2\overline{R}_0^4}{(3+4a\overline{R}_0^2+a^2\overline{R}_0^4)^2} \,. \label{K-qG-SD}
\end{eqnarray}

\subsection{Diffractive excitation of a projectile gluon}

Turning now to the diffractive gluon excitations, the differential SD cross sections can be written as
\begin{eqnarray} 
\nonumber
\frac{d\sigma^{G\to q\bar q}_{\rm SD}}{d\Omega} &\simeq& \frac{{\cal K}^{G\to q\bar q}_{\rm SD}(s,\hat{s},\alpha)}{(2\pi)^2}\,g(x_g,\mu^2)\,
\int d^2\rho d^2\rho'\, e^{i\vec \kappa (\vec \rho-\vec \rho\,')}\,(\vec{\rho}\cdot \vec{\rho}\,') \\
&\times& \overline{\sum}\hat{\Psi}_{G\to q\bar q}(\vec \rho,\alpha)
\hat{\Psi}^{\dagger}_{G\to q\bar q}(\vec \rho\,',\alpha)\,, \qquad \hat{s}=x_g\,s  \,, \label{SD-CS-qq-fin} \\
\nonumber
\frac{d\sigma^{G\to G_1G_2}_{\rm SD}}{d\Omega} &\simeq& \frac{{\cal K}^{G\to G_1G_2}_{\rm SD}(s,\hat{s},\alpha)}{(2\pi)^2}\,g(x_g,\mu^2)\,
\int d^2\rho d^2\rho'\, e^{i\vec \kappa (\vec \rho-\vec \rho\,')}\,(\vec{\rho}\cdot \vec{\rho}\,') \\
&\times& \overline{\sum}\hat{\Psi}_{G\to G_1G_2}(\vec \rho,\alpha)\hat{\Psi}^{\dagger}_{G\to G_1G_2}(\vec \rho\,',\alpha) \,,
\end{eqnarray}
for $G\to q\bar q$ and $G\to G_1G_2$ subprocesses, respectively. In analogy with the diffractive bremsstrahlung process 
discussed in detail above, we find
\begin{eqnarray}
{\cal K}^{G\to q\bar q}_{\rm SD}&=&\frac{1}{B_{\rm SD}}\,\frac{9a\overline{\sigma}_0(\hat{s})^2}{256\pi}
\Big\{ {\cal W}_1(\hat{s})\,\Big[ \frac{16}{27} - \frac{4\alpha}{3} + \alpha^2 \Big] + 
{\cal W}_2(\hat{s})\,\Big[ \frac{32}{27} - \frac{8\alpha}{3} + \alpha^2 \Big] \Big\} \,, \label{KGqq} \\
{\cal K}^{G\to G_1G_2}_{\rm SD}&=&\frac{1}{B_{\rm SD}}\,\frac{9a\overline{\sigma}_0(\hat{s})^2}{256\pi}
\Big\{ {\cal W}_1(\hat{s})\,\Big[ \frac{5}{6} - \alpha\bar \alpha \Big] + 
{\cal W}_2(\hat{s})\,\Big[ \frac{1}{6} - \alpha\bar \alpha \Big] \Big\} \,, \label{KGGG}
\end{eqnarray}
where ${\cal W}_{1,2}$ are defined above in Eq.~(\ref{K-qG-SD}). In what follows, these formulas will be used 
in analysis of the SD-to-inclusive ratio.

\section{Diffractive to inclusive ratio}
\label{Sec:SD-to-incl}
The CDF Run II experimental data \cite{Aaltonen:2012tha} on SD dijet production are given, in particular, in terms of the SD-to-inclusive 
ratio ${\cal R}_{\rm SD/incl}$, which is defined as follows
\begin{eqnarray}
\label{R-SD-incl}
{\cal R}_{\rm SD/incl}=\frac{\Delta \sigma_{\rm SD}/\Delta \xi}{\Delta \sigma_{\rm incl}}\,, \qquad \Delta \xi = 0.06 \,, \qquad 
\xi \equiv 1-x_F = \frac{M_X^2}{s} \,,
\end{eqnarray}
where $M_X$ is the invariant mass squared of the diffractive system $X$, $M_X^2$, containing the dijet, $x_F$ is the Feynman variable of the recoil antiproton,
$\Delta \sigma_{\rm SD}$ ($\Delta \sigma_{\rm incl}$) are the SD (inclusive) dijet cross sections integrated over the detector acceptance 
regions in $\xi \equiv 1-x_F$ variable, $0.03<\xi<0.09$, in jet pseudorapidities, $|\eta_{1,2}|<2.5$, in jet transverse energies, $E_T^{1,2}>5$ GeV, 
and in the antiproton transverse momentum squared, $|t|<1$ GeV$^2$. The SD-to-inclusive ratio is then measured as function of the hard scale 
$Q^2\gg R_0^2$ of the dijet and $x_{\rm Bj}$,
\begin{eqnarray}
\label{Q2-xBj}
Q^2=\frac{(E_T^{1}+E_T^{2})^2}{4} \,, \qquad x_{\rm Bj} = \frac{1}{\sqrt{s}}\sum_{i=1}^{3\,{\rm jets}}E_T^i e^{-\eta_i}\,.
\end{eqnarray}

It is difficult to make one-to-one correspondence between theory and data for the observables entering  Eq.~(\ref{R-SD-incl}), but one can rely on 
approximations. Considering, for example, the gluon Bremsstrahlung mechanism $q\to qG$ as a suitable example which was thoroughly discussed in the previous 
sections, a dominant contribution to the sum in Eq.~(\ref{Q2-xBj}) comes from the high-$p_T$ gluon jet $G$ with a small longitudinal momentum
fraction $x_G\ll 1$. Indeed, in the high-$p_T$ limit, the leading jets are mostly back-to-back, i.e. $p_T^G \sim p_T^q \sim E^{1,2}_T$, 
the third subleading jet is more likely to be produced at a smaller transverse momentum $p_T^{{\rm jet}=3}\ll E^{1,2}_T$, while the gluon 
Bremstrahlung is enhanced at small $\alpha\ll 1$ and thus is radiated at smallest pseudorapidity among the leading jets such that
\begin{eqnarray}
x_G = x_q \alpha \,, \qquad x_G \ll x_q < 1 \,.
\end{eqnarray}
Besides, the invariant mass squared of the dijet system, $M^2$, can be approximately identified with the hard scale $Q^2$, i.e.
\begin{eqnarray}
\mu^2\simeq M^2 \simeq Q^2 \,.
\end{eqnarray}
As we will see below, these approximations are vital for a comparison of the dipole model results with the data.

In the experimental definition (\ref{R-SD-incl}), the numerator
\begin{eqnarray}
\label{numer}
\frac{\Delta \sigma_{\rm SD}}{\Delta \xi} \sim \frac{d\sigma_{\rm SD}}{d\xi}
\end{eqnarray}
is essentially the differential SD dijet cross section averaged over the bin interval $\Delta \xi$.
The dipole formula for the differential SD dijet cross section (\ref{SD-CS-fin})
is differential in dijet mass squared $M^2=Q^2$ (or $x_2$), and not in $M_X^2$, so the analysis of its $\xi$ dependence
as well as implementation of $\xi$ cuts cannot be directly performed. Following the proposal of Ref.~\cite{Pasechnik:2012ac}, 
the way out of this issue is to employ the $\xi$-dependence provided by the phenomenological SD cross section in 
the triple-Regge form \cite{Kazarinov:1975kw}
\begin{eqnarray}
- \frac{d^2\sigma_{\rm SD}^{\rm pp}}{d\xi\,dq_{\perp}^2} = \sqrt{\frac{s_1}{s}}\, \frac{G_{\Pom\Pom\Reg}(0)}{\xi^{3/2}}\,
e^{-B^{\rm pp}_{\Pom\Pom\Reg}\,q_{\perp}^2} + \frac{G_{3\Pom}(0)}{\xi}\, e^{-B^{\rm pp}_{3\Pom}\,q_{\perp}^2}\,,
\label{triple-R}
\end{eqnarray}
such that the main effect of constraints on $\xi$ variable in this Regge-based cross section and in our result (\ref{SD-CS-fin}) 
is expected to be roughly the same. In the above formula (\ref{triple-R}), we use the results of Ref.~\cite{Kazarinov:1975kw}
\begin{eqnarray*}
&& s_1=1\,\GeV^2\,, \qquad B^{\rm pp}_{\Pom\Pom i} = R^2_{\Pom\Pom i} - 2\alpha^\prime_\Pom\,\ln\xi \,, \qquad i=\Pom,\,\Reg \,, \\
&& G_{3\Pom}(0) = G_{\Pom\Pom\Reg}(0)= 3.2\,\mb/\GeV^2 \,, \quad R^2_{3\Pom}=4.2\,\GeV^{-2}\,, \quad R^2_{\Pom\Pom\Reg}=1.7\,\GeV^{-2} \,,
\end{eqnarray*}
where $\alpha^\prime_\Pom\approx 0.25\GeV^{-2}$ is the Pomeron trajectory slope. Although these parameters were determined by the fit 
to data long time ago at relatively low energies (ISR), they well predicted data on diffraction at LHC (see Appendix A in \cite{pi-pi}). 
\begin{figure*}[!h]
 \centerline{\includegraphics[width=0.75\textwidth]{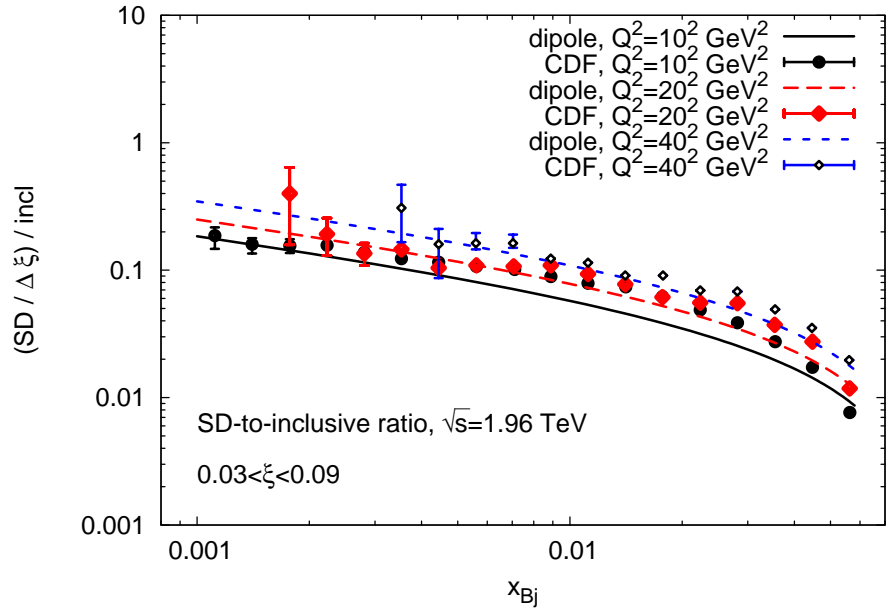}}
   \caption{
   \small The SD-to-inclusive ratio ${\cal R}_{\rm SD/incl}(x_{\rm Bj},Q^2)$ given by Eq.~(\ref{R-SD-incl-fin}) 
   as function of $x_{\rm Bj}$ for three different values of the hard scale $Q^2=10^2,\,20^2$ and $40^2$ GeV$^2$ 
   in comparison to the CDF Run II data \cite{Aaltonen:2012tha}.}
 \label{fig:SD-to-incl}
\end{figure*}

When integrating Eq.~(\ref{triple-R}) over $\xi$ interval allowed by the detector constraints, its upper limit is equal to the maximal measured 
$\xi_{\rm max}=0.09$ (the largest momentum that can be taken by the ``active'' gluon) while its minimal value coincides 
with $x_{\rm Bj}$ characterising the hard dijet system. Then, the correction factor relating the integrated SD cross section 
with the experimentally constrained $\Delta \sigma_{\rm SD}$ as a function of $x_{\rm Bj}$ reads
\begin{eqnarray}
\delta=\frac{\int dt\, \int_{x_{\rm Bj}}^{\xi_{\rm max}} d\xi\,\frac{d^2\sigma}{dt d\xi}}
{\int dt\int_{\xi^*}^{0.3} d\xi\,\frac{d^2\sigma}{dt d\xi}} \,, \qquad \xi^*=\frac{Q^2}{s}\ll x_{\rm Bj} \,,
\label{delta-cut}
\end{eqnarray}
where $\xi^*$ is associated with the minimal produced diffractive mass $X$, containing only the dijet. As the result is practically 
non-sensitive to the upper limit of $\xi$, we fix it to $0.3$ corresponding to a situation when a constituent quark in the target looses 
most of its energy into a hard radiation of the $t$-channel gluon \cite{Pasechnik:2012ac}. Notably, the correction factor (\ref{delta-cut}) 
automatically accounts for the jet pseudorapidity constraint such that the resulting SD cross section vanishes when approaching 
the kinematical boundary $x_{\rm Bj}\to \xi_{\rm max}$ as expected.
\begin{figure*}[!h]
 \centerline{\includegraphics[width=0.75\textwidth]{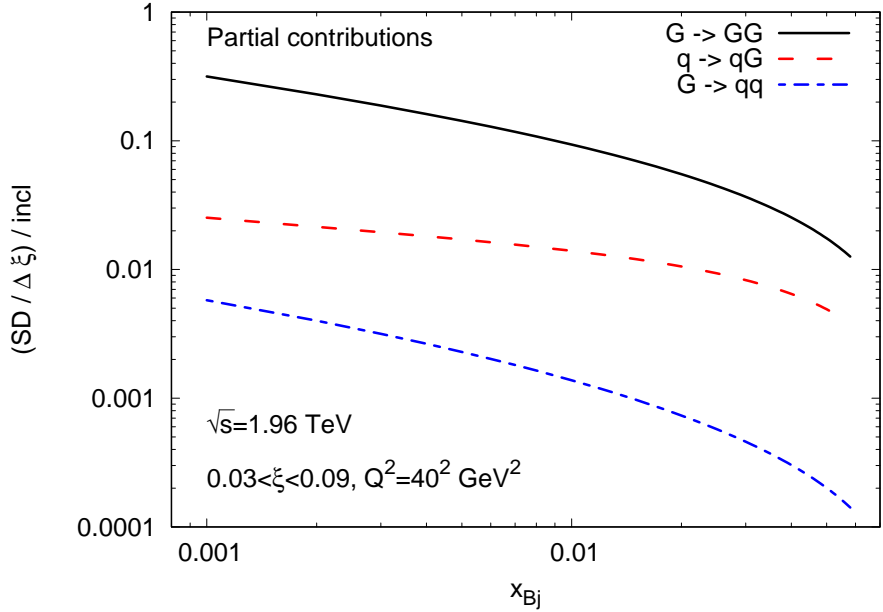}}
   \caption{
   \small The SD-to-inclusive ratio ${\cal R}_{\rm SD/incl}(x_{\rm Bj},Q^2)$ given by Eq.~(\ref{R-SD-incl-fin}) where 
   $\sigma_{\rm SD}$ corresponds to one of the three partial contributions $G\to GG$ (solid line), $q\to qG$ (dashed line) 
   and $G\to q\bar q$ (dash-dotted line)
   as functions of $x_{\rm Bj}$ for a fixed value of the hard scale $Q^2=40^2$ GeV$^2$ and at Tevatron energy $\sqrt{s}=1.96$ TeV.}
 \label{fig:SD-to-incl-partial}
\end{figure*}

The most important results of the previous sections, are the dipole formulas for the differential inclusive and SD dijet cross sections given 
by Eqs.~(\ref{inclCS}) and (\ref{SD-CS-fin}), respectively. One immediately notices that the differential cross sections (\ref{inclCS}) and 
(\ref{SD-CS-fin}) are proportional to each other, similarly to what was seen earlier in the case of Abelian radiation in Refs.~\cite{Pasechnik:2011nw,
Pasechnik:2012ac,Pasechnik:2014lga}. When calculating the SD-to-inclusive ratio, however, one notices that ${\cal K}_{\rm SD}$ and
${\cal K}_{\rm incl}$ are functions of $\alpha$ which has to be integrated out in the corresponding cross sections. For example, using the above 
results with $q\to q+G$ subprocess, we obtain
\begin{eqnarray} \nonumber
\Delta\sigma^{q\to qG}_{\rm SD}&\equiv& \frac{d\sigma^{q\to qG}_{\rm SD}}{dx_G} = \delta \int_{x_G}^1 \frac{d\alpha}{\alpha^2}\,
{\cal K}^{q\to qG}_{\rm SD}\Big(s,\hat{s},\alpha\Big)\sum_{q,\bar{q}}\Big[ q\Big(x_q,Q^2\Big)+
\bar{q}\Big(x_q,Q^2\Big) \Big] \\
&\times& \frac{4}{3}\frac{\alpha_s(Q^2)}{\pi}\int_{\rho_{\rm min}}^{\rho_{\rm max}} d\rho\, \rho^3\,
\Big\{m_q^2\alpha^4\,K_0^2(\tau\rho)+\Big[1+(1-\alpha)^2\Big]\tau^2\,K_1^2(\tau\rho)\Big\} \,,
\label{CS-fin-SD}
\end{eqnarray}
where 
\begin{eqnarray}
x_q = \frac{x_G}{\alpha} \,, \qquad \hat{s} = s\,x_q \,,
\end{eqnarray}
and $\tau=\tau(\alpha)$ is defined in Eq.~(\ref{psi-qG}), ${\cal K}^{q\to qG}_{\rm SD}$ is defined in Eq.~(\ref{K-qG-SD}), 
and the integration limits are $\rho_{\rm min}\sim 1/Q$ and $\rho_{\rm min}\sim 
1/E^{1,2}_{T,{\rm min}}$, $E^{1,2}_{T,{\rm min}}=5$ GeV. 
\begin{figure*}[!h]
 \centerline{\includegraphics[width=0.75\textwidth]{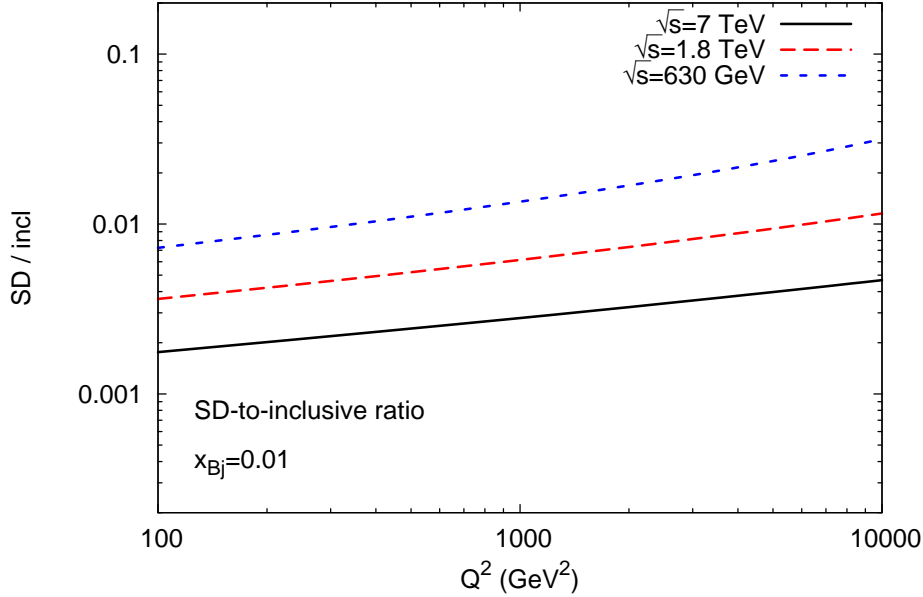}}
   \caption{
   \small The SD-to-inclusive ratio ${\cal R}_{\rm SD/incl}(x_{\rm Bj},Q^2)$ 
   as function of $Q^2$ for three different values of the c.m. energy $\sqrt{s}=630$ GeV, $1.8$ TeV and $7$ TeV. 
   No additional phase-space correction factor $\delta$ and no division by $\Delta\xi$ have been applied here.}
 \label{fig:SD-to-incl-Q2}
\end{figure*}

Analogically, for the inclusive dijet cross section for the gluon Bremsstrahlung $q\to qG$ subprocess, we write
\begin{eqnarray} \nonumber
\Delta\sigma^{q\to qG}_{\rm incl}&\equiv& \frac{d\sigma^{q\to qG}_{\rm incl}}{dx_G} = \int_{x_G}^1 \frac{d\alpha}{\alpha^2}\,
{\cal K}^{q\to qG}_{\rm incl}\Big(x_2,\alpha\Big)\sum_{q,\bar{q}}\Big[ q\Big(x_q,Q^2\Big)+
\bar{q}\Big(x_q,Q^2\Big) \Big] \\
&\times& \frac{4}{3}\frac{\alpha_s(Q^2)}{\pi}\int_{\rho_{\rm min}}^{\rho_{\rm max}} d\rho\, \rho^3\,
\Big\{m_q^2\alpha^4\,K_0^2(\tau\rho)+\Big[1+(1-\alpha)^2\Big]\tau^2\,K_1^2(\tau\rho)\Big\} \,,
\label{CS-fin-INC}
\end{eqnarray}
where ${\cal K}^{q\to qG}_{\rm incl}$ is defined in Eq.~(\ref{x2}), and $x_2 = Q^2/sx_G$ (see also Ref.~\cite{Pasechnik:2011nw}). The cross sections for 
the gluon-initiated subprocesses, such as $G\to q\bar q$ and $G\to G_1G_2$, can be obtained in complete analogy to the above expressions, 
except that the (anti)quark densities are replaced by the gluon one.

Finally, the SD-to-inclusive ratio is written as follows
\begin{eqnarray}
\label{R-SD-incl-fin}
{\cal R}_{\rm SD/incl}=\frac{1}{\Delta \xi}\,\frac{d\sigma^{q\to qG}_{\rm SD}/dx_G+d\sigma^{G\to q\bar q}_{\rm SD}/dx_G+
d\sigma^{G\to G_1G_2}_{\rm SD}/dx_G}
{d\sigma^{q\to qG}_{\rm incl}/dx_G+d\sigma^{G\to q\bar q}_{\rm incl}/dx_G+d\sigma^{G\to G_1G_2}_{\rm incl}/dx_G}\,,
\end{eqnarray}
accounting for the proper phase space constraints. In Fig.~\ref{fig:SD-to-incl} we show the SD-to-inclusive ratio
${\cal R}_{\rm SD/incl}$ computed by using Eq.~(\ref{R-SD-incl-fin}) as function of $x_{\rm Bj}$ variable 
for three different values of the hard scale $Q^2=10^2,\,20^2$ and $40^2$ GeV$^2$ and compared to the corresponding 
CDF Run II data \cite{Aaltonen:2012tha}. In addition, in Fig.~\ref{fig:SD-to-incl-partial} we show partial contributions 
to the SD-to-inclusive ratio ${\cal R}_{\rm SD/incl}(x_{\rm Bj},Q^2)$ at fixed $Q^2=40^2$ GeV$^2$ corresponding to 
$G\to GG$ (solid line), $q\to qG$ (dashed line) and $G\to q\bar q$ (dash-dotted line) subprocesses. Apparently, $G\to GG$
process is dominant in the SD production of di-jets in the considered kinematical region.

Notice that while the GBW parametrization of the dipole cross section, Eq.~(\ref{ansatz}), 
is sufficiently accurate for many applications, the DGLAP evolution within the used scale range might be not negligible. Therefore, 
we introduced here a scale dependence of the parameter $R_0$ in (\ref{ansatz}) in accordance with the model
\cite{bartels}. We found the effect rather small, but it somewhat improves agreement with data.

The energy and hard scale dependences of the SD-to-inclusive ratio ${\cal R}_{\rm SD/incl}$ are typically considered to be 
an important qualitative measure of the diffractive factorisation breaking. Similarly to the SD Drell-Yan \cite{Kopeliovich:2006tk,Pasechnik:2011nw} and gauge 
boson \cite{Pasechnik:2012ac} production cases, an important feature of the ratio ${\cal R}_{\rm SD/incl}$ in the SD dijet production case
also inconsistent with a factorisation-based analysis, is its unusual energy and scale dependence shown in Fig.~\ref{fig:SD-to-incl-Q2}.
It appears to be remarkably universal for both SD Abelian and non-Abelian types of radiation. As was discussed earlier in 
Refs.~\cite{Pasechnik:2015fxa,Kopeliovich:2016rts}, the ratio ${\cal R}_{\rm SD/incl}$, in particular, its normalisation and slopes 
in $\sqrt{s}$ and $Q^2$, is sensitive only to a particular (process-dependent) linear combination of the universal dipole cross section 
evaluated at different separations causing an interplay between hard and soft fluctuations (see also Ref.~\cite{Kopeliovich:2006tk}). 
Notably, the sign of these slopes is the same for all the SD reactions, that have been studied in the dipole picture so far, but it is 
clearly opposite to that in the existing factorisation-based predictions (c.f.~Ref.~\cite{Luszczak:2017pna}). In this sense, 
the SD-to-inclusive ratio can be used as an important probe for the QCD mechanism of diffraction that is essentially determined 
by an interplay between hard and soft interactions. As a possible direction for future studies, in order to quantify the factorisation 
breaking effects, it would be instructive to make a more detailed comparison between the predictions of the dipole and factorisation-based models.

\section{Summary and conclusions}
\label{Sec:summary}

In this work, we  computed the inclusive and single-diffractive cross sections for dijet production in hadron-hadron collisions 
in the dipole picture accounting for the quark ($q\to qG$) and gluon ($G\to q\bar q,\, GG$) excitations.
Applied for the kinematics of the CDF experiment at the Tevatron, we estimated the  SD-to-inclusive cross section ratio 
${\cal R}_{\rm SD/incl}$ as function of $x_{\rm Bj}$ and 
hard scale of the process $Q^2$. Diffractive factorisation is found to be severely broken 
for many reasons.  

First, the diffractive structure functions, measured in the diagonal diffractive DIS, should not be applied to an off-diagonal hadronic 
diffraction, like dijet production. Such a mismatch causes dramatic effects, usually related to the rapidity gap survival probability. 
Working at the amplitude level in the dipole representation the gap survival factor is by default embedded into our calculation 
of the diffractive amplitude. 

The gap survival amplitude is related to possible interactions  with the spectator partons. However, we found that factorisation is broken 
even in the case of diffractive excitation of a projectile quark, $q\to qG$, the process free of spectators. Our calculations within the dipole formalism
results in a cross section falling with relative jet transverse momentum as $1/\kappa^6$ , 
while the factorisation would lead to $1/\kappa^4$ dependence.

Remarkably, interactions with the spectator partons in the projectile hadron, not only suppress the cross section, but also considerably 
increase it, giving rise to a new mechanism of diffractive dijet production. Interaction with the spectator quarks, separated by large transverse 
distance from the active one, causes an interplay of the long-range interactions with the spectator partons, and the
hard-scale interactions with a given Fock state. A similar conclusion, which has resulted in a dynamically calculated rapidity 
gap survival factor derived from the modelling of multiparton interactions, has been made in Ref.~\cite{Rasmussen:2015qgr}.

The results for ${\cal R}_{\rm SD/incl}(x_{\rm Bj},Q^2)$ exhibit an overall consistency with the data available from Tevatron. Notice that
these results for non-Abelian (gluon Bremsstrahlung and splitting) types of radiation,
and SD Abelian diffractive radiation (Drell-Yan \cite{Kopeliovich:2006tk,Pasechnik:2011nw}) demonstrate 
an interesting similarity in shapes and magnitudes, pointing at a universal character of the diffractive factorisation 
breaking effects in hadronic diffraction.

\vspace{0.5cm}

{\bf Acknowledgments} 
B.K. and I.P. are partially supported by Fondecyt grant No. 1170319 (Chile), by Proyecto Basal FB 0821 (Chile),
and by Conicyt grant  PIA ACT1406 (Chile). R.P. is partially supported by the Swedish Research Council, contract 
numbers 621-2013-4287 and 2016-05996, by CONICYT grants PIA ACT1406 and MEC80170112, as well as by 
the European Research Council (ERC) under the European Union's Horizon 2020 research and innovation programme 
(grant agreement No 668679). This work was supported in part by the Ministry of Education, Youth and Sports 
of the Czech Republic, project LT17018. The work has been performed in the framework of COST Action CA15213 
``Theory of hot matter and relativistic heavy-ion collisions'' (THOR).


\end{document}